\input harvmac
\def\ev#1{\langle#1\rangle}
\def\sign{{\rm sign}}

\input amssym
\input epsf.tex

\newcount\figno
\figno=0
\def\fig#1#2#3{
\par\begingroup\parindent=0pt\leftskip=1cm\rightskip=1cm\parindent=0pt
\baselineskip=11pt
\global\advance\figno by 1
\midinsert
\epsfxsize=#3
\centerline{\epsfbox{#2}}
\vskip 12pt
{\bf Fig. \the\figno:} #1\par
\endinsert\endgroup\par
}
\def\figlabel#1{\xdef#1{\the\figno}}
\def\encadremath#1{\vbox{\hrule\hbox{\vrule\kern8pt\vbox{\kern8pt
\hbox{$\displaystyle #1$}\kern8pt}
\kern8pt\vrule}\hrule}}


\newfam\frakfam
\font\teneufm=eufm10
\font\seveneufm=eufm7
\font\fiveeufm=eufm5
\textfont\frakfam=\teneufm
\scriptfont\frakfam=\seveneufm
\scriptscriptfont\frakfam=\fiveeufm


\def\bb{
\font\tenmsb=msbm10
\font\sevenmsb=msbm7
\font\fivemsb=msbm5
\textfont1=\tenmsb
\scriptfont1=\sevenmsb
\scriptscriptfont1=\fivemsb
}



\newfam\dsromfam
\font\tendsrom=dsrom10
\textfont\dsromfam=\tendsrom
\def\ds{\fam\dsromfam \tendsrom}


\newfam\mbffam
\font\tenmbf=cmmib10
\font\sevenmbf=cmmib7
\font\fivembf=cmmib5
\textfont\mbffam=\tenmbf
\scriptfont\mbffam=\sevenmbf
\scriptscriptfont\mbffam=\fivembf


\newfam\mbfcalfam
\font\tenmbfcal=cmbsy10
\font\sevenmbfcal=cmbsy7
\font\fivembfcal=cmbsy5
\textfont\mbfcalfam=\tenmbfcal
\scriptfont\mbfcalfam=\sevenmbfcal
\scriptscriptfont\mbfcalfam=\fivembfcal


\newfam\mscrfam
\font\tenmscr=rsfs10
\font\sevenmscr=rsfs7
\font\fivemscr=rsfs5
\textfont\mscrfam=\tenmscr
\scriptfont\mscrfam=\sevenmscr
\scriptscriptfont\mscrfam=\fivemscr




\def\tilde{\widetilde}

\def\bar{\overline}
\def\b{\bar}
\def\bsq#1{{{\b{#1}}^{\lower 2.5pt\hbox{$\scriptstyle 2$}}}}
\def\bexp#1#2{{{\b{#1}}^{\lower 2.5pt\hbox{$\scriptstyle #2$}}}}
\def\dotexp#1#2{{{#1}^{\lower 2.5pt\hbox{$\scriptstyle #2$}}}}


\def\rt2{\sqrt{2}}
\def\half {{1 \over 2}}

\def\Tr{\mathop{\rm Tr}}

\def\sign{\mathop{\rm sign}}


\font\tenbifull=cmmib10
\font\tenbimed=cmmib7
\font\tenbismall=cmmib5
\textfont9=\tenbifull \scriptfont9=\tenbimed
\scriptscriptfont9=\tenbismall

\mathchardef\bbGamma="7000
\mathchardef\bbDelta="7001
\mathchardef\bbPhi="7002
\mathchardef\bbAlpha="7003
\mathchardef\bbXi="7004
\mathchardef\bbPi="7005
\mathchardef\bbSigma="7006
\mathchardef\bbUpsilon="7007
\mathchardef\bbTheta="7008
\mathchardef\bbPsi="7009
\mathchardef\bbOmega="700A
\mathchardef\bbalpha="710B
\mathchardef\bbbeta="710C
\mathchardef\bbgamma="710D
\mathchardef\bbdelta="710E
\mathchardef\bbepsilon="710F
\mathchardef\bbzeta="7110
\mathchardef\bbeta="7111
\mathchardef\bbtheta="7112
\mathchardef\bbiota="7113
\mathchardef\bbkappa="7114
\mathchardef\bblambda="7115
\mathchardef\bbmu="7116
\mathchardef\bbnu="7117
\mathchardef\bbxi="7118
\mathchardef\bbpi="7119
\mathchardef\bbrho="711A
\mathchardef\bbsigma="711B
\mathchardef\bbtau="711C
\mathchardef\bbupsilon="711D
\mathchardef\bbphi="711E
\mathchardef\bbchi="711F
\mathchardef\bbpsi="7120
\mathchardef\bbomega="7121
\mathchardef\bbvarepsilon="7122
\mathchardef\bbvartheta="7123
\mathchardef\bbvarpi="7124
\mathchardef\bbvarrho="7125
\mathchardef\bbvarsigma="7126
\mathchardef\bbvarphi="7127





\def\CL{{\cal L}}

\def\CN{{\cal N}}
\def\CO{{\cal O}}


\def\1{{\ds 1}}

\def\Z{\hbox{$\bb Z$}}


\noblackbox

\def\unit{\relax{\rm 1\kern-.26em I}}
\def\nada{\relax{\rm 0\kern-.30em l}}
\def\tilde{\widetilde}


\noblackbox
\def\IL{\relax{\rm I\kern-.18em L}}
\def\IH{\relax{\rm I\kern-.18em H}}
\def\IR{\relax{\rm I\kern-.18em R}}
\def\IC{\relax\hbox{$\inbar\kern-.3em{\rm C}$}}
\def\IZ{\relax\ifmmode\mathchoice
{\hbox{\cmss Z\kern-.4em Z}}{\hbox{\cmss Z\kern-.4em Z}} {\lower.9pt\hbox{\cmsss Z\kern-.4em Z}}
{\lower1.2pt\hbox{\cmsss Z\kern-.4em Z}}\else{\cmss Z\kern-.4em Z}\fi}

\def\CN {{\cal N}}

\def\partialslash{\not{\hbox{\kern-2pt $\partial$}}}

\def\CL {{\cal L}}

\def\CO {{\cal O}}


\def\CN {{\cal N}}

\def\CO {{\cal O}}

\def\Tr{{\rm Tr}}

\font\manual=manfnt \def\dbend{\lower3.5pt\hbox{\manual\char127}}

\def\IZ{\relax\ifmmode\mathchoice
{\hbox{\cmss Z\kern-.4em Z}}{\hbox{\cmss Z\kern-.4em Z}} {\lower.9pt\hbox{\cmsss Z\kern-.4em Z}}
{\lower1.2pt\hbox{\cmsss Z\kern-.4em Z}}\else{\cmss Z\kern-.4em Z}\fi}
\def\half {{1\over 2}}

\def\lfm#1{\medskip\noindent\item{#1}}

\def\bar{\overline}

\def\rt2{\sqrt{2}}
\def\irt2{{1\over\sqrt{2}}}

\def\slashchar#1{\setbox0=\hbox{$#1$}           
   \dimen0=\wd0                                 
   \setbox1=\hbox{/} \dimen1=\wd1               
   \ifdim\dimen0>\dimen1                        
      \rlap{\hbox to \dimen0{\hfil/\hfil}}      
      #1                                        
   \else                                        
      \rlap{\hbox to \dimen1{\hfil$#1$\hfil}}   
      /                                         
   \fi}

\def\foursqr#1#2{{\vcenter{\vbox{
    \hrule height.#2pt
    \hbox{\vrule width.#2pt height#1pt \kern#1pt
    \vrule width.#2pt}
    \hrule height.#2pt
    \hrule height.#2pt
    \hbox{\vrule width.#2pt height#1pt \kern#1pt
    \vrule width.#2pt}
    \hrule height.#2pt
        \hrule height.#2pt
    \hbox{\vrule width.#2pt height#1pt \kern#1pt
    \vrule width.#2pt}
    \hrule height.#2pt
        \hrule height.#2pt
    \hbox{\vrule width.#2pt height#1pt \kern#1pt
    \vrule width.#2pt}
    \hrule height.#2pt}}}}
\def\psqr#1#2{{\vcenter{\vbox{\hrule height.#2pt
    \hbox{\vrule width.#2pt height#1pt \kern#1pt
    \vrule width.#2pt}
    \hrule height.#2pt \hrule height.#2pt
    \hbox{\vrule width.#2pt height#1pt \kern#1pt
    \vrule width.#2pt}
    \hrule height.#2pt}}}}
\def\sqr#1#2{{\vcenter{\vbox{\hrule height.#2pt
    \hbox{\vrule width.#2pt height#1pt \kern#1pt
    \vrule width.#2pt}
    \hrule height.#2pt}}}}

\def\figin{\epsfcheck\figin}\def\figins{\epsfcheck\figins}
\def\epsfcheck{\ifx\epsfbox\UnDeFiNeD
\message{(NO epsf.tex, FIGURES WILL BE IGNORED)}
\gdef\figin##1{\vskip2in}\gdef\figins##1{\hskip.5in}
\else\message{(FIGURES WILL BE INCLUDED)}%
\gdef\figin##1{##1}\gdef\figins##1{##1}\fi}
\def\DefWarn#1{}
\def\figinsert{\goodbreak\midinsert}
\def\ifig#1#2#3{\DefWarn#1\xdef#1{fig.~\the\figno}
\writedef{#1\leftbracket fig.\noexpand~\the\figno}%
\figinsert\figin{\centerline{#3}}\medskip\centerline{\vbox{\baselineskip12pt \advance\hsize by
-1truein\noindent\footnotefont{\bf Fig.~\the\figno:\ } \it#2}}
\bigskip\endinsert\global\advance\figno by1}
\lref\WittenHF{
  E.~Witten,
  ``Quantum Field Theory and the Jones Polynomial,''
Commun.\ Math.\ Phys.\  {\bf 121}, 351 (1989).
}
\lref\AharonyBX{
  O.~Aharony, A.~Hanany, K.~A.~Intriligator, N.~Seiberg and M.~J.~Strassler,
  ``Aspects of N = 2 supersymmetric gauge theories in three dimensions,''
  Nucl.\ Phys.\  B {\bf 499}, 67 (1997)
  [arXiv:hep-th/9703110].
}

\lref\HitchinEA{
  N.~J.~Hitchin, A.~Karlhede, U.~Lindstrom and M.~Rocek,
  ``Hyperkahler Metrics and Supersymmetry,''
Commun.\ Math.\ Phys.\  {\bf 108}, 535 (1987).
}

\lref\AganagicUW{
  M.~Aganagic, K.~Hori, A.~Karch and D.~Tong,
  ``Mirror symmetry in (2+1)-dimensions and (1+1)-dimensions,''
JHEP {\bf 0107}, 022 (2001).
[hep-th/0105075].
}

\lref\AffleckAS{
  I.~Affleck, J.~A.~Harvey and E.~Witten,
  ``Instantons and (Super)Symmetry Breaking in (2+1)-Dimensions,''
Nucl.\ Phys.\ B {\bf 206}, 413 (1982).
}
\lref\ClossetVG{
  C.~Closset, T.~T.~Dumitrescu, G.~Festuccia, Z.~Komargodski and N.~Seiberg,
  ``Contact Terms, Unitarity, and F-Maximization in Three-Dimensional Superconformal Theories,''
JHEP {\bf 1210}, 053 (2012).
[arXiv:1205.4142 [hep-th]].
}
\lref\taurr{
  E.~Barnes, E.~Gorbatov, K.~A.~Intriligator, M.~Sudano and J.~Wright,
  ``The Exact superconformal R-symmetry minimizes tau(RR),''
Nucl.\ Phys.\ B {\bf 730}, 210 (2005).
[hep-th/0507137].
}
\lref\CalliasKG{
  C.~Callias,
  ``Index Theorems on Open Spaces,''
Commun.\ Math.\ Phys.\  {\bf 62}, 213 (1978).
}
\lref\OikonomouTSA{
  V.~K.~Oikonomou,
  ``Supersymmetric Chern-Simons vortex systems and extended supersymmetric quantum mechanics algebras,''
Nucl.\ Phys.\ B {\bf 870}, 477 (2013).
[arXiv:1308.0461 [hep-th]].
}
\lref\NekrasovUH{
  N.~A.~Nekrasov and S.~L.~Shatashvili,
  ``Supersymmetric vacua and Bethe ansatz,''
Nucl.\ Phys.\ Proc.\ Suppl.\  {\bf 192-193}, 91 (2009).
[arXiv:0901.4744 [hep-th]].
}
\lref\IntriligatorEX{
  K.~A.~Intriligator and N.~Seiberg,
  ``Mirror symmetry in three-dimensional gauge theories,''
Phys.\ Lett.\ B {\bf 387}, 513 (1996).
[hep-th/9607207].
}
\lref\GatesNR{
  S.~J.~Gates, M.~T.~Grisaru, M.~Rocek and W.~Siegel,
  ``Superspace Or One Thousand and One Lessons in Supersymmetry,''
Front.\ Phys.\  {\bf 58}, 1 (1983).
[hep-th/0108200].
}
\lref\PolyakovFU{
  A.~M.~Polyakov,
  ``Quark Confinement and Topology of Gauge Groups,''
Nucl.\ Phys.\ B {\bf 120}, 429 (1977).
}
\lref\AganagicUW{
  M.~Aganagic, K.~Hori, A.~Karch and D.~Tong,
  ``Mirror symmetry in (2+1)-dimensions and (1+1)-dimensions,''
JHEP {\bf 0107}, 022 (2001).
[hep-th/0105075].
}
\lref\IntriligatorCP{
  K.~A.~Intriligator and N.~Seiberg,
  ``Lectures on Supersymmetry Breaking,''
Class.\ Quant.\ Grav.\  {\bf 24}, S741 (2007).
[hep-ph/0702069].
}
\lref\IvanovFN{
  E.~A.~Ivanov,
  ``Chern-Simons matter systems with manifest N=2 supersymmetry,''
Phys.\ Lett.\ B {\bf 268}, 203 (1991).
}
\lref\IntriligatorRX{
  K.~A.~Intriligator, N.~Seiberg and S.~H.~Shenker,
  ``Proposal for a simple model of dynamical SUSY breaking,''
Phys.\ Lett.\ B {\bf 342}, 152 (1995).
[hep-ph/9410203].
}
\lref\ClossetVP{
  C.~Closset, T.~T.~Dumitrescu, G.~Festuccia, Z.~Komargodski and N.~Seiberg,
  ``Comments on Chern-Simons Contact Terms in Three Dimensions,''
[arXiv:1206.5218 [hep-th]].
}
\lref\DoreyKQ{
  N.~Dorey, D.~Tong and S.~Vandoren,
  ``Instanton effects in three-dimensional supersymmetric gauge theories with matter,''
JHEP {\bf 9804}, 005 (1998).
[hep-th/9803065].
}
\lref\KapustinPK{
  A.~Kapustin, E.~Witten and ,
  ``Electric-Magnetic Duality And The Geometric Langlands Program,''
Commun.\ Num.\ Theor.\ Phys.\  {\bf 1}, 1 (2007).
[hep-th/0604151].
}
\lref\KimUW{
  C.~-j.~Kim,
  ``Selfdual vortices in the generalized Abelian Higgs model with independent Chern-Simons interaction,''
Phys.\ Rev.\ D {\bf 47}, 673 (1993).
[hep-th/9209110].
}
\lref\SchroersZY{
  B.~J.~Schroers,
  ``The Spectrum of Bogomol'nyi solitons in gauged linear sigma models,''
Nucl.\ Phys.\ B {\bf 475}, 440 (1996).
[hep-th/9603101].
}

\lref\PeninSI{
  A.~A.~Penin, V.~A.~Rubakov, P.~G.~Tinyakov and S.~V.~Troitsky,
  ``What becomes of vortices in theories with flat directions,''
Phys.\ Lett.\ B {\bf 389}, 13 (1996).
[hep-ph/9609257].
}
\lref\AchucarroII{
  A.~Achucarro, A.~C.~Davis, M.~Pickles and J.~Urrestilla,
  ``Vortices in theories with flat directions,''
Phys.\ Rev.\ D {\bf 66}, 105013 (2002).
[hep-th/0109097].
}
\lref\BuchbinderEM{
  I.~L.~Buchbinder, N.~G.~Pletnev and I.~B.~Samsonov,
  ``Effective action of three-dimensional extended supersymmetric matter on gauge superfield background,''
JHEP {\bf 1004}, 124 (2010).
[arXiv:1003.4806 [hep-th]].
}
\lref\HitchinEA{
  N.~J.~Hitchin, A.~Karlhede, U.~Lindstrom and M.~Rocek,
  ``Hyperkahler Metrics and Supersymmetry,''
Commun.\ Math.\ Phys.\  {\bf 108}, 535 (1987).
}
\lref\AganagicUW{
  M.~Aganagic, K.~Hori, A.~Karch and D.~Tong,
  ``Mirror symmetry in (2+1)-dimensions and (1+1)-dimensions,''
JHEP {\bf 0107}, 022 (2001).
[hep-th/0105075].
}
\lref\JafferisNS{
  D.~Jafferis and X.~Yin,
  ``A Duality Appetizer,''
[arXiv:1103.5700 [hep-th]].
}
\lref\BashkirovVY{
  D.~Bashkirov,
  ``Aharony duality and monopole operators in three dimensions,''
[arXiv:1106.4110 [hep-th]].
}
\lref\KlebanovTD{
  I.~R.~Klebanov, S.~S.~Pufu, S.~Sachdev and B.~R.~Safdi,
  ``Entanglement Entropy of 3-d Conformal Gauge Theories with Many Flavors,''
JHEP {\bf 1205}, 036 (2012).
[arXiv:1112.5342 [hep-th]].
}
\lref\KomargodskiPC{
  Z.~Komargodski and N.~Seiberg,
  ``Comments on the Fayet-Iliopoulos Term in Field Theory and Supergravity,''
JHEP {\bf 0906}, 007 (2009).
[arXiv:0904.1159 [hep-th]].
}
\lref\IStwo{K. Intriligator and N. Seiberg, {\bf will discuss it here}}
\lref\HwangJH{
  C.~Hwang, H.~-C.~Kim and J.~Park,
  ``Factorization of the 3d superconformal index,''
[arXiv:1211.6023 [hep-th]].
}
\lref\WittenBS{
  E.~Witten,
  ``Toroidal compactification without vector structure,''
JHEP {\bf 9802}, 006 (1998).
[hep-th/9712028].
}
\lref\WilczekDU{
  F.~Wilczek,
  ``Magnetic Flux, Angular Momentum, and Statistics,''
Phys.\ Rev.\ Lett.\  {\bf 48}, 1144 (1982).
}
\lref\WeinbergI{
  E.~J.~Weinberg,
  ``Index Calculations for the Fermion-Vortex System,''
Phys.\ Rev.\ D {\bf 24}, 2669 (1981).
}
\lref\SeibergAJ{
  N.~Seiberg and E.~Witten,
  ``Monopoles, duality and chiral symmetry breaking in N=2 supersymmetric QCD,''
Nucl.\ Phys.\ B {\bf 431}, 484 (1994).
[hep-th/9408099].
}
\lref\CollieMX{
  B.~Collie and D.~Tong,
  ``The Dynamics of Chern-Simons Vortices,''
Phys.\ Rev.\ D {\bf 78}, 065013 (2008).
[arXiv:0805.0602 [hep-th]].
}
\lref\KimYZ{
  Y.~Kim and K.~-M.~Lee,
  ``Vortex dynamics in selfdual Chern-Simons Higgs systems,''
Phys.\ Rev.\ D {\bf 49}, 2041 (1994).
[hep-th/9211035].
}
\lref\WittenYC{
  E.~Witten,
  ``Phases of N=2 theories in two-dimensions,''
Nucl.\ Phys.\ B {\bf 403}, 159 (1993).
[hep-th/9301042].
}
\lref\SmilgaUY{
  A.~V.~Smilga,
  ``Once more on the Witten index of 3d supersymmetric YM-CS theory,''
JHEP {\bf 1205}, 103 (2012).
[arXiv:1202.6566 [hep-th]].
}
\lref\HenningsonVB{
  M.~Henningson,
  ``Ground states of supersymmetric Yang-Mills-Chern-Simons theory,''
JHEP {\bf 1211}, 013 (2012).
[arXiv:1209.1798 [hep-th]].
}
\lref\AcharyaDZ{
  B.~S.~Acharya and C.~Vafa,
  ``On domain walls of N=1 supersymmetric Yang-Mills in four-dimensions,''
[hep-th/0103011].
}
\lref\JafferisUN{
  D.~L.~Jafferis,
  ``The Exact Superconformal R-Symmetry Extremizes Z,''
JHEP {\bf 1205}, 159 (2012).
[arXiv:1012.3210 [hep-th]].
}
\lref\ClossetRU{
  C.~Closset, T.~T.~Dumitrescu, G.~Festuccia and Z.~Komargodski,
  ``Supersymmetric Field Theories on Three-Manifolds,''
[arXiv:1212.3388 [hep-th]].
}
\lref\BergmanNA{
  O.~Bergman, A.~Hanany, A.~Karch and B.~Kol,
  ``Branes and supersymmetry breaking in three-dimensional gauge theories,''
JHEP {\bf 9910}, 036 (1999).
[hep-th/9908075].
}
\lref\PoppitzKZ{
  E.~Poppitz and M.~Unsal,
  ``Chiral gauge dynamics and dynamical supersymmetry breaking,''
JHEP {\bf 0907}, 060 (2009).
[arXiv:0905.0634 [hep-th]].
}
\lref\OhtaIV{
  K.~Ohta,
  ``Supersymmetric index and s rule for type IIB branes,''
JHEP {\bf 9910}, 006 (1999).
[hep-th/9908120].
}
\lref\SeibergPQ{
  N.~Seiberg,
  ``Electric - magnetic duality in supersymmetric nonAbelian gauge theories,''
Nucl.\ Phys.\ B {\bf 435}, 129 (1995).
[hep-th/9411149].
}
\lref\ARSW{
  O.~Aharony, S.~S.~Razamat, N.~Seiberg and B.~Willett,
  ``3d dualities from 4d dualities,''
[arXiv:1305.3924 [hep-th]].
}
\lref\PasquettiFJ{
  S.~Pasquetti,
  ``Factorisation of N = 2 Theories on the Squashed 3-Sphere,''
JHEP {\bf 1204}, 120 (2012).
[arXiv:1111.6905 [hep-th]].
}
\lref\BeemMB{
  C.~Beem, T.~Dimofte and S.~Pasquetti,
  ``Holomorphic Blocks in Three Dimensions,''
[arXiv:1211.1986 [hep-th]].
}
\lref\IntriligatorID{
  K.~A.~Intriligator and N.~Seiberg,
  ``Duality, monopoles, dyons, confinement and oblique confinement in supersymmetric SO(N(c)) gauge theories,''
Nucl.\ Phys.\ B {\bf 444}, 125 (1995).
[hep-th/9503179].
}
\lref\DineFI{M. Dine, ``Fields, Strings, and Duality: TASI 96," eds. C. Efthimiou and B. Greene, (World Scientific, Singapore, 1997).}
\lref\CveticXN{
  M.~Cvetic, T.~W.~Grimm and D.~Klevers,
  ``Anomaly Cancellation And Abelian Gauge Symmetries In F-theory,''
JHEP {\bf 1302}, 101 (2013).
[arXiv:1210.6034 [hep-th]].
}
\lref\WittenDS{
  E.~Witten,
  ``Supersymmetric index of three-dimensional gauge theory,''
In *Shifman, M.A. (ed.): The many faces of the superworld* 156-184.
[hep-th/9903005].
}
\lref\KatzTH{
  S.~H.~Katz and C.~Vafa,
  ``Geometric engineering of N=1 quantum field theories,''
Nucl.\ Phys.\ B {\bf 497}, 196 (1997).
[hep-th/9611090].
}
\lref\AffleckAS{
  I.~Affleck, J.~A.~Harvey and E.~Witten,
  ``Instantons and (Super)Symmetry Breaking in (2+1)-Dimensions,''
Nucl.\ Phys.\ B {\bf 206}, 413 (1982).
}
\lref\StrasslerHY{
  M.~J.~Strassler,
  ``Confining phase of three-dimensional supersymmetric quantum electrodynamics,''
In *Shifman, M.A. (ed.): The many faces of the superworld* 262-279.
[hep-th/9912142].
}
\lref\BorokhovIB{
  V.~Borokhov, A.~Kapustin and X.~-k.~Wu,
  ``Topological disorder operators in three-dimensional conformal field theory,''
JHEP {\bf 0211}, 049 (2002).
[hep-th/0206054].
}
\lref\TongQD{
  D.~Tong,
  ``Quantum Vortex Strings: A Review,''
Annals Phys.\  {\bf 324}, 30 (2009).
[arXiv:0809.5060 [hep-th]].
}

\lref\MaldacenaSS{
  J.~M.~Maldacena, G.~W.~Moore and N.~Seiberg,
  ``D-brane charges in five-brane backgrounds,''
JHEP {\bf 0110}, 005 (2001).
[hep-th/0108152].
}
\lref\PantevRH{
  T.~Pantev and E.~Sharpe,
  ``Notes on gauging noneffective group actions,''
  arXiv:hep-th/0502027.
}
\lref\PantevZS{
  T.~Pantev and E.~Sharpe,
  ``GLSM's for gerbes (and other toric stacks),''
  Adv.\ Theor.\ Math.\ Phys.\  {\bf 10}, 77 (2006)
  [arXiv:hep-th/0502053].
}
\lref\CaldararuTC{
  A.~Caldararu, J.~Distler, S.~Hellerman, T.~Pantev and E.~Sharpe,
  ``Non-birational twisted derived equivalences in abelian GLSMs,''
  arXiv:0709.3855 [hep-th].
}
\lref\SeibergQD{
  N.~Seiberg,
  ``Modifying the Sum Over Topological Sectors and Constraints on Supergravity,''
JHEP {\bf 1007}, 070 (2010).
[arXiv:1005.0002 [hep-th]].
}
\lref\BanksZN{
  T.~Banks and N.~Seiberg,
  ``Symmetries and Strings in Field Theory and Gravity,''
Phys.\ Rev.\ D {\bf 83}, 084019 (2011).
[arXiv:1011.5120 [hep-th]].
}
\lref\HellermanFV{
  S.~Hellerman and E.~Sharpe,
  ``Sums over topological sectors and quantization of Fayet-Iliopoulos parameters,''
Adv.\ Theor.\ Math.\ Phys.\  {\bf 15}, 1141 (2011).
[arXiv:1012.5999 [hep-th]].
}

\lref\SeibergNZ{
  N.~Seiberg and E.~Witten,
  ``Gauge dynamics and compactification to three-dimensions,''
In *Saclay 1996, The mathematical beauty of physics* 333-366.
[hep-th/9607163].
}
\lref\CollieIZ{
  B.~Collie and D.~Tong,
  ``The Partonic Nature of Instantons,''
JHEP {\bf 0908}, 006 (2009).
[arXiv:0905.2267 [hep-th]].
}
\lref\DunneQY{
  G.~V.~Dunne,
  ``Aspects of Chern-Simons theory,''
[hep-th/9902115].
}
\lref\OlmezAU{
  S.~Olmez and M.~Shifman,
  ``Revisiting Critical Vortices in Three-Dimensional SQED,''
Phys.\ Rev.\ D {\bf 78}, 125021 (2008).
[arXiv:0808.1859 [hep-th]].
}
\lref\Shamirthesis{
 I.~Shamir,
 ``Aspects of three dimensional Seiberg duality,''
 M. Sc. thesis submitted to the Weizmann Institute of Science, April 2010.
 }
\lref\ClossetVP{
  C.~Closset, T.~T.~Dumitrescu, G.~Festuccia, Z.~Komargodski and N.~Seiberg,
  ``Comments on Chern-Simons Contact Terms in Three Dimensions,''
[arXiv:1206.5218 [hep-th]].
}
\lref\KaoGF{
  H.~-C.~Kao, K.~-M.~Lee and T.~Lee,
  ``The Chern-Simons coefficient in supersymmetric Yang-Mills Chern-Simons theories,''
Phys.\ Lett.\ B {\bf 373}, 94 (1996).
[hep-th/9506170].
}

\lref\KoroteevRB{
  P.~Koroteev, M.~Shifman, W.~Vinci and A.~Yung,
  ``Quantum Dynamics of Low-Energy Theory on Semilocal Non-Abelian Strings,''
Phys.\ Rev.\ D {\bf 84}, 065018 (2011).
[arXiv:1107.3779 [hep-th]].
}

\lref\FestucciaWS{
  G.~Festuccia and N.~Seiberg,
  ``Rigid Supersymmetric Theories in Curved Superspace,''
JHEP {\bf 1106}, 114 (2011).
[arXiv:1105.0689 [hep-th]].
}

\lref\FendleyVE{
  P.~Fendley and K.~A.~Intriligator,
  ``Scattering and thermodynamics of fractionally charged supersymmetric solitons,''
Nucl.\ Phys.\ B {\bf 372}, 533 (1992).
[hep-th/9111014].
}
\lref\FendleyDM{
  P.~Fendley and K.~A.~Intriligator,
  ``Scattering and thermodynamics in integrable N=2 theories,''
Nucl.\ Phys.\ B {\bf 380}, 265 (1992).
[hep-th/9202011].
}
\lref\GatesQN{
  S.~J.~Gates, Jr. and H.~Nishino,
  ``Remarks on the N=2 supersymmetric Chern-Simons theories,''
Phys.\ Lett.\ B {\bf 281}, 72 (1992).
}
\lref\ZupnikRY{
  B.~M.~Zupnik and D.~G.~Pak,
  ``Topologically Massive Gauge Theories In Superspace,''
Sov.\ Phys.\ J.\  {\bf 31}, 962 (1988).
}
\lref\IntriligatorAU{
  K.~A.~Intriligator and N.~Seiberg,
  ``Lectures on supersymmetric gauge theories and electric - magnetic duality,''
Nucl.\ Phys.\ Proc.\ Suppl.\  {\bf 45BC}, 1 (1996).
[hep-th/9509066].
}
\lref\WittenDF{
  E.~Witten,
  ``Constraints on Supersymmetry Breaking,''
Nucl.\ Phys.\ B {\bf 202}, 253 (1982).
}
\lref\WilczekCY{
  F.~Wilczek and A.~Zee,
  ``Linking Numbers, Spin, and Statistics of Solitons,''
Phys.\ Rev.\ Lett.\  {\bf 51}, 2250 (1983).
}
\lref\WilczekDU{
  F.~Wilczek,
  ``Magnetic Flux, Angular Momentum, and Statistics,''
Phys.\ Rev.\ Lett.\  {\bf 48}, 1144 (1982).
}
\lref\EdalatiVK{
  M.~Edalati and D.~Tong,
  ``Heterotic Vortex Strings,''
JHEP {\bf 0705}, 005 (2007).
[hep-th/0703045 [HEP-TH]].
}
\lref\MezincescuGB{
  L.~Mezincescu and P.~K.~Townsend,
  ``Semionic Supersymmetric Solitons,''
J.\ Phys.\ A A {\bf 43}, 465401 (2010).
[arXiv:1008.2775 [hep-th]].
}
\lref\TaubesTM{
  C.~H.~Taubes,
  ``Arbitrary N: Vortex Solutions to the First Order Landau-Ginzburg Equations,''
Commun.\ Math.\ Phys.\  {\bf 72}, 277 (1980).
}
\lref\MorrisonFR{
  D.~R.~Morrison and M.~R.~Plesser,
  ``Summing the instantons: Quantum cohomology and mirror symmetry in toric varieties,''
Nucl.\ Phys.\ B {\bf 440}, 279 (1995).
[hep-th/9412236].
}
\lref\WessCP{
  J.~Wess and J.~Bagger,
  ``Supersymmetry and supergravity,''
Princeton, USA: Univ. Pr. (1992) 259 p.
}
\lref\deBoerKR{
  J.~de Boer, K.~Hori and Y.~Oz,
  ``Dynamics of N=2 supersymmetric gauge theories in three-dimensions,''
Nucl.\ Phys.\ B {\bf 500}, 163 (1997).
[hep-th/9703100].
}
\lref\GoldhaberKN{
  A.~S.~Goldhaber, A.~Rebhan, P.~van Nieuwenhuizen and R.~Wimmer,
  ``Quantum corrections to mass and central charge of supersymmetric solitons,''
Phys.\ Rept.\  {\bf 398}, 179 (2004).
[hep-th/0401152].
}
\lref\DumitrescuIU{
  T.~T.~Dumitrescu and N.~Seiberg,
  ``Supercurrents and Brane Currents in Diverse Dimensions,''
JHEP {\bf 1107}, 095 (2011).
[arXiv:1106.0031 [hep-th]].
}
\lref\LeePM{
  B.~-H.~Lee and H.~Min,
  ``Quantum aspects of supersymmetric Maxwell Chern-Simons solitons,''
Phys.\ Rev.\ D {\bf 51}, 4458 (1995).
[hep-th/9409006].
}
\lref\GiveonZN{
  A.~Giveon and D.~Kutasov,
  ``Seiberg Duality in Chern-Simons Theory,''
Nucl.\ Phys.\ B {\bf 812}, 1 (2009).
[arXiv:0808.0360 [hep-th]].
}
\lref\AharonyGP{
  O.~Aharony,
  ``IR duality in d = 3 N=2 supersymmetric USp(2N(c)) and U(N(c)) gauge theories,''
Phys.\ Lett.\ B {\bf 404}, 71 (1997).
[hep-th/9703215].
}
\lref\LeeEQ{
  C.~-k.~Lee, K.~-M.~Lee and H.~Min,
  ``Selfdual Maxwell Chern-Simons solitons,''
Phys.\ Lett.\ B {\bf 252}, 79 (1990).
}
\lref\ShifmanCE{
  M.~Shifman and A.~Yung,
  ``Supersymmetric Solitons and How They Help Us Understand Non-Abelian Gauge Theories,''
Rev.\ Mod.\ Phys.\  {\bf 79}, 1139 (2007).
[hep-th/0703267].
}
\lref\HananyEA{
  A.~Hanany and D.~Tong,
  ``Vortex strings and four-dimensional gauge dynamics,''
JHEP {\bf 0404}, 066 (2004).
[hep-th/0403158].
}
\lref\HananyHP{
  A.~Hanany and D.~Tong,
  ``Vortices, instantons and branes,''
JHEP {\bf 0307}, 037 (2003).
[hep-th/0306150].
}
\lref\KapustinHPK{
  A.~Kapustin and B.~Willett,
  ``Wilson loops in supersymmetric Chern-Simons-matter theories and duality,''
[arXiv:1302.2164 [hep-th]].
}
\lref\IntriligatorPU{
  K.~A.~Intriligator and S.~D.~Thomas,
  ``Dynamical supersymmetry breaking on quantum moduli spaces,''
Nucl.\ Phys.\ B {\bf 473}, 121 (1996).
[hep-th/9603158].
}
\lref\IzawaPK{
  K.~-I.~Izawa and T.~Yanagida,
  ``Dynamical supersymmetry breaking in vector - like gauge theories,''
Prog.\ Theor.\ Phys.\  {\bf 95}, 829 (1996).
[hep-th/9602180].
}
\lref\BeemMB{
  C.~Beem, T.~Dimofte and S.~Pasquetti,
  ``Holomorphic Blocks in Three Dimensions,''
[arXiv:1211.1986 [hep-th]].
}
\lref\ShifmanDR{
  M.~Shifman and A.~Yung,
  ``NonAbelian string junctions as confined monopoles,''
Phys.\ Rev.\ D {\bf 70}, 045004 (2004).
[hep-th/0403149].
}

\lref\AharonyHDA{
  O.~Aharony, N.~Seiberg and Y.~Tachikawa,
  ``Reading between the lines of four-dimensional gauge theories,''
[arXiv:1305.0318 [hep-th]].
}
\lref\KimQMA{
  Y.~-b.~Kim and K.~-M.~Lee,
  ``First and second order vortex dynamics,''
Phys.\ Rev.\ D {\bf 66}, 045016 (2002).
[hep-th/0204111].
}
\lref\AuzziFS{
  R.~Auzzi, S.~Bolognesi, J.~Evslin, K.~Konishi and A.~Yung,
Nucl.\ Phys.\ B {\bf 673}, 187 (2003).
[hep-th/0307287].
}
\lref\SamolsNE{
  T.~M.~Samols,
  ``Vortex scattering,''
Commun.\ Math.\ Phys.\  {\bf 145}, 149 (1992)..
}
\lref\WittenNV{
  E.~Witten,
  ``Supersymmetric index in four-dimensional gauge theories,''
Adv.\ Theor.\ Math.\ Phys.\  {\bf 5}, 841 (2002).
[hep-th/0006010].
}

\lref\LeeYC{
  B.~-H.~Lee, C.~-k.~Lee and H.~Min,
  ``Supersymmetric Chern-Simons vortex systems and fermion zero modes,''
Phys.\ Rev.\ D {\bf 45}, 4588 (1992).
}
\lref\WillettGP{
  B.~Willett and I.~Yaakov,
  ``N=2 Dualities and Z Extremization in Three Dimensions,''
[arXiv:1104.0487 [hep-th]].
}
\lref\WardIJ{
  R.~S.~Ward,
  ``Slowly Moving Lumps In The Cp**1 Model In (2+1)-dimensions,''
Phys.\ Lett.\ B {\bf 158}, 424 (1985).
}
\lref\KacGW{
  V.~G.~Kac and A.~V.~Smilga,
  ``Vacuum structure in supersymmetric Yang-Mills theories with any gauge group,''
In *Shifman, M.A. (ed.): The many faces of the superworld* 185-234.
[hep-th/9902029].
}
\lref\WittenNV{
  E.~Witten,
  ``Supersymmetric index in four-dimensional gauge theories,''
Adv.\ Theor.\ Math.\ Phys.\  {\bf 5}, 841 (2002).
[hep-th/0006010].
}
\lref\JackiwPR{
  R.~Jackiw, K.~-M.~Lee and E.~J.~Weinberg,
  ``Selfdual Chern-Simons solitons,''
Phys.\ Rev.\ D {\bf 42}, 3488 (1990).
}
\lref\JackiwAW{
  R.~Jackiw and E.~J.~Weinberg,
  ``Selfdual Chern-simons Vortices,''
Phys.\ Rev.\ Lett.\  {\bf 64}, 2234 (1990).
}
\lref\WeinbergAW{
  E.~J.~Weinberg,
  ``Multivortex Solutions of the Ginzburg-landau Equations,''
Phys.\ Rev.\ D {\bf 19}, 3008 (1979).
}
\lref\HindmarshYY{
  M.~Hindmarsh,
  ``Semilocal topological defects,''
Nucl.\ Phys.\ B {\bf 392}, 461 (1993).
[hep-ph/9206229].
}
\lref\VachaspatiDZ{
  T.~Vachaspati and A.~Achucarro,
  ``Semilocal cosmic strings,''
Phys.\ Rev.\ D {\bf 44}, 3067 (1991).
}
\lref\PasquettiFJ{
  S.~Pasquetti,
  ``Factorisation of N = 2 Theories on the Squashed 3-Sphere,''
JHEP {\bf 1204}, 120 (2012).
[arXiv:1111.6905 [hep-th]].
}
\lref\OlmezAU{
  S.~Olmez and M.~Shifman,
  ``Revisiting Critical Vortices in Three-Dimensional SQED,''
Phys.\ Rev.\ D {\bf 78}, 125021 (2008).
[arXiv:0808.1859 [hep-th]].
}
\lref\LeeseFN{
  R.~A.~Leese and T.~M.~Samols,
  ``Interaction of semilocal vortices,''
Nucl.\ Phys.\ B {\bf 396}, 639 (1993).
}
\lref\IntriligatorAN{
  K.~A.~Intriligator,
  ``Fusion residues,''
Mod.\ Phys.\ Lett.\ A {\bf 6}, 3543 (1991).
[hep-th/9108005].
}
\lref\WittenXI{
  E.~Witten,
  ``The Verlinde algebra and the cohomology of the Grassmannian,''
In *Cambridge 1993, Geometry, topology, and physics* 357-422.
[hep-th/9312104].
}
\lref\HananyVM{
  A.~Hanany and K.~Hori,
  ``Branes and N=2 theories in two-dimensions,''
Nucl.\ Phys.\ B {\bf 513}, 119 (1998).
[hep-th/9707192].
}
\lref\DoreyRB{
  N.~Dorey and D.~Tong,
  ``Mirror symmetry and toric geometry in three-dimensional gauge theories,''
JHEP {\bf 0005}, 018 (2000).
[hep-th/9911094].
}
\lref\PisarskiYJ{
  R.~D.~Pisarski and S.~Rao,
  ``Topologically Massive Chromodynamics in the Perturbative Regime,''
Phys.\ Rev.\ D {\bf 32}, 2081 (1985).
}
\lref\TongKY{
  D.~Tong,
  ``Dynamics of N=2 supersymmetric Chern-Simons theories,''
JHEP {\bf 0007}, 019 (2000).
[hep-th/0005186].
}
\lref\GaiottoQI{
  D.~Gaiotto and X.~Yin,
  ``Notes on superconformal Chern-Simons-Matter theories,''
JHEP {\bf 0708}, 056 (2007).
[arXiv:0704.3740 [hep-th]].
}
\lref\GukovSN{
  S.~Gukov, E.~Witten and ,
  ``Rigid Surface Operators,''
Adv.\ Theor.\ Math.\ Phys.\  {\bf 14} (2010).
[arXiv:0804.1561 [hep-th]].
}
\lref\DimofteJU{
  T.~Dimofte, D.~Gaiotto and S.~Gukov,
  ``Gauge Theories Labelled by Three-Manifolds,''
[arXiv:1108.4389 [hep-th]].
}
\lref\BorokhovCG{
  V.~Borokhov, A.~Kapustin and X.~-k.~Wu,
  ``Monopole operators and mirror symmetry in three-dimensions,''
JHEP {\bf 0212}, 044 (2002).
[hep-th/0207074].
}
\lref\BeniniMF{
  F.~Benini, C.~Closset and S.~Cremonesi,
  ``Comments on 3d Seiberg-like dualities,''
  JHEP {\bf 1110}, 075 (2011)
  [arXiv:1108.5373 [hep-th]].
}
\lref\DyerFJA{
  E.~Dyer, M‡r.~Mezei and S.~S.~Pufu,
  ``Monopole Taxonomy in Three-Dimensional Conformal Field Theories,''
[arXiv:1309.1160 [hep-th]].
}
\lref\KapustinHA{
  A.~Kapustin and M.~J.~Strassler,
  ``On mirror symmetry in three-dimensional Abelian gauge theories,''
JHEP {\bf 9904}, 021 (1999).
[hep-th/9902033].
}
\lref\ElitzurNR{
  S.~Elitzur, G.~W.~Moore, A.~Schwimmer and N.~Seiberg,
  ``Remarks on the Canonical Quantization of the Chern-Simons-Witten Theory,''
Nucl.\ Phys.\ B {\bf 326}, 108 (1989).
}
\lref\IntriligatorUE{
  K.~Intriligator, H.~Jockers, P.~Mayr, D.~R.~Morrison and M.~R.~Plesser,
  ``Conifold Transitions in M-theory on Calabi-Yau Fourfolds with Background Fluxes,''
[arXiv:1203.6662 [hep-th]].
}
\lref\aspects{
  K.~Intriligator and N.~Seiberg,
  ``Aspects of 3d N=2 Chern-Simons-Matter Theories,''
[arXiv:1305.1633 [hep-th]].
}
\lref\AharonyDHA{
  O.~Aharony, S.~S.~Razamat, N.~Seiberg and B.~Willett,
  ``3d dualities from 4d dualities,''
JHEP {\bf 1307}, 149 (2013).
[arXiv:1305.3924 [hep-th]].
}
\lref\AharonyKMA{
  O.~Aharony, S.~S.~Razamat, N.~Seiberg and B.~Willett,
  ``3$d$ dualities from 4$d$ dualities for orthogonal groups,''
JHEP {\bf 1308}, 099 (2013).
[arXiv:1307.0511 [hep-th]].
}

\newbox\tmpbox\setbox\tmpbox\hbox{\abstractfont }
\Title{\vbox{\baselineskip12pt \hbox{UCSD-PTH-14-02}}}
{\vbox{\centerline{Matching 3d N=2 Vortices and Monopole Operators}}}
\smallskip
\centerline{Kenneth Intriligator}
\smallskip
\bigskip
\centerline{{\it Department of Physics, University of
California, San Diego, La Jolla, CA 92093 USA}}

\bigskip
\vskip 1cm

\noindent In earlier work with N. Seiberg, we explored connections between monopole operators, the Coulomb branch modulus, and vortices for 3d, $\CN =2$ supersymmetric, $U(1)_k$  Chern-Simons matter theories. We here extend the monopole / vortex matching analysis,  to theories with general matter electric charges.  We verify, for general matter content, that the spin and other quantum numbers of the chiral monopole operators match those of corresponding BPS vortex states, at the top and bottom of the tower associated with quantizing the vortices' Fermion zero modes. There are associated subtleties from non-normalizable Fermi zero modes, which contribute non-trivially to the BPS vortex spectrum and monopole operator matching; a proposed interpretation is further discussed here.

\bigskip

\Date{June 2014}

\newsec{Introduction}

Three-dimensional $U(1)$ gauge theories exhibit IR-interesting phenomena and phases, with qualitative similarities to 4d non-Abelian gauge theories.  For example, electric-magnetic dualities can be explored in this context, and the $U(1)$  gauge group makes it easier to make the duality more precise, and potentially construct the duality-map between fields.  This is particularly true for 3d theories with $\CN\geq 2$ supersymmetry, where magnetically charged, BPS vortex solitons can be regarded as giving the dual quanta in terms of the electric variables,
with  corresponding chiral superfield monopole operators.

Building on \aspects, we here consider 3d, $\CN =2$ supersymmetric, compact\foot{I.e. gauge transformations are $A_\mu \to A_\mu +\partial _\mu f$, with $f\sim f+2\pi$, which requires $n_i\in \Z$ and $q_J\in \Z$.  However, $U(1)\not\subset SU(2)$, so there is no instanton sum.  The monopole operators here are of singular, Dirac-type, with unobservable string thanks to the quantization conditions. } $U(1)_k$ gauge theory ($k$ is the Chern-Simons coefficient), with matter chiral superfields $Q_i$, with general electric charges $n_i\in \Z$.     A key aspect is that the theory has an exact\foot{In other theories, $U(1)_J$ can be explicitly broken by short-distance physics, which can add monopole operators to ${\cal L}_{eff}$; then $U(1)_J$ is at best an accidental, approximate symmetry, if those operators are irrelevant, or a fine-tuning if they are not.  E.g. if $U(1)\subset SU(2)$, instantons in the UV $SU(2)/U(1)$ explicitly break $U(1)_J$ \refs{\PolyakovFU,\AffleckAS}.   In our susy context, the monopole operators are chiral superfields, and holomorphy constrains their possible   appearance in the superpotential.}, conserved  global $U(1)_J$ topological symmetry, with current $j_J^\mu = \epsilon ^{\mu \rho \sigma}F_{\rho \sigma}/4\pi$, and associated charge
\eqn\qjis{U(1)_J: \qquad\qquad q_J=\int {F_{12}\over 2\pi}\in \Z.}  
The theory contains local operators, and particle states, with $q_J\neq 0$, despite the fact that the 
photon and $Q_i$ have $q_J=0$.  There are three distinct, related ways to get $q_J\neq 0$:
\lfm{1.} Monopole operators: disorder the gauge field, with $q_J$ units of magnetic flux, around a point $x_0^\mu$ in spacetime \refs{\KapustinHA, \BorokhovIB, \BorokhovCG}. It is a local, chiral $\CN =2$ operator (the 3d reduction of 4d 't Hooft line operators).  This short-distance definition of the operator is independent of IR data, e.g. 
the particular vacua, or the spacetime geometry.  The chiral condition implies that  the real scalar $\sigma = \Sigma |$ of the $\CN =2$ photon linear multiplet has  \refs{\BorokhovCG, \aspects}
\eqn\sigmar{\sigma (x)\to {q_J\over r_{3d}}, \qquad  \hbox{where}\qquad r_{3d}\equiv ||x^\mu -x_0^\mu ||_{\rm Euclidean}.}

\lfm{2.} On the $\sigma \neq 0$ Coulomb branch, if it exists, $U(1)_J$  is spontaneously broken and the associated, compact NG boson, $a\sim a+2\pi$, can be identified with the dualized photon \AffleckAS.  The gauge field linear multiplet $\Sigma =-{i\over 2}\bar D DV$ can be dualized to chiral superfields  \refs{\HitchinEA, \deBoerKR}, and exponentiated to obtain chiral operators  \AharonyBX\ with $U(1)_J$ charge $q_J=\pm 1$:
 \eqn\xis{X_{\pm}\sim e^{\pm (2\pi \sigma/e_{eff}^2  +ia)}.}
The microscopic, monopole disorder operator of the theory at the origin is also denoted as $X_\pm$, with  \xis\ its low-energy effective description.   The $U(1)_J$ charge $q_J$ chiral operator is $X_+^{q_J}$ for $q_J>0$, or $X_-^{|q_J|}$ for $q_J<0$.  The $X_+$ or $X_-$ monopole operator is only $U(1)_{gauge}$ neutral if the corresponding Coulomb branch exists. 

  \lfm{3.} BPS vortex particle field configurations exist in certain Higgs vacua,  $\ev{Q_i}\neq 0$, when the FI parameter $\zeta \neq 0$. Their BPS mass is $m=|Z|=|\zeta q_J|$.  Using $z=x+iy$ for the 2d spatial plane, the gauge field $A_z\equiv \half (A_x-iA_y)$ and matter wind at infinity as
  \eqn\awind{A_z\to {q_J\over 2iz}+\dots,\qquad \hbox{for}\qquad |z|\to \infty}
 \eqn\qwind{Q_i\to  e^{-i n_i q_J\theta}\left(Q_i^{vac}+{\rho _i \over |z|}+\dots \right)\qquad\hbox{for}\qquad |z|\to \infty.} 
 \noindent Upon taking $\zeta \to 0$, all $Q^{vac}_i\to 0$, the BPS magnetic vortices become massless, and can  potentially condense and
give dual Higgs description of the Coulomb branch\AharonyBX, in the sense of 3d mirror symmetry's exchange of the electric and magnetic Higgs and Coulomb branches \IntriligatorEX.     See also  \refs{\PasquettiFJ, \BeemMB} for vortices and partition functions.

Connections and distinctions between monopole operators, vortices, and the Coulomb branch, for the theories in flat space, were studied in  \refs{ \aspects, \AharonyKMA}, and will be further explored here. We determine, and match, the gauge and global charges of monopole operators and the vortices.   For the 
monopole operators $X_\pm$, the charges  are simply, and exactly, obtained by a one-loop calculation of induced Chern-Simons terms \refs{\AharonyBX,\ClossetVG, \ClossetVP, \aspects} to be 
 \eqn\onengen{
\vbox{\offinterlineskip\tabskip=0pt
\halign{\strut\vrule#
&~$#$~\hfil\vrule
&~$#$~\hfil\vrule
&~$#$~\hfil\vrule
&~$#$~\hfil\vrule
&~$#$~\hfil\vrule
&~$#$~\hfil\vrule
&~$#$~\hfil
&\vrule#
\cr
\noalign{\hrule}
& & U(1)_{\rm gauge} & U(1)_j & U(1)_R &  U(1)_J \cr
\noalign{\hrule}
&  Q_i        &\; n_i & \;\delta _{ij} &\;0& \quad 0   \cr
&X_\pm  &\;  -(k_c\pm k)&\; -\half  |n_j|  &\; \half \sum _i |n_i|  & \; \pm 1 \cr
}\hrule}}
with $k_c\equiv \half \sum _i n_i |n_i|$ (see sect. 2).    The operators $X_\pm $ in \onengen\
exist as  gauge invariant operators\foot{The  superconformal $U(1)_{R_*}$ of the $\CN =2$ SCFT at  $Q_i=X_\pm =0$, is a linear combination of those in \onengen, $U(1)_{R_*}=U(1)_R+\sum _j R_j U(1)_j$, so $\Delta (Q_i)=R_i$, and $\Delta (X_\pm)=R(X_\pm)=\half \sum _i |n_i|(1-R _i)$, with $R_i$ determined by F-extremization \JafferisUN (or $\tau _{RR}$ minimization \refs{\taurr,\ClossetVG}).} only if $k=\mp k_c$; this is the condition for the $\ev{X_\pm}$ Coulomb branch to exist.

The corresponding charges of BPS vortices arise in a seemingly different way, from quantizing the  vortex Fermion zero modes\foot{Vs on $S^2\times \IR$, where the $\sigma \neq 0$ in \sigmar\ lifts all of the monopole operator's Fermi zero modes \BorokhovCG. 
This fits with the radial quantization map between energy on  $S^2\times \IR$ and operator dimension.}, $\Psi _A$, with $A=1\dots N_z$, i.e. from
 \eqn\fermiquant{\{\Psi _A, \Psi _B^\dagger\}=\delta _{AB}, \qquad A, B=1\dots N_z.}
This formally gives a tower of $2^{N_z}$ degenerate states: treating the $\Psi _A$ ($\Psi _A^\dagger$) as raising (lowering) operators,  the top and bottom vortex states in this tower are 
 \eqn\omegapm{|\Omega _\pm\rangle _{q_J}, \qquad\hbox{with} \qquad \Psi _A^\dagger |\Omega _+\rangle _{q_J}= \Psi _A |\Omega _-\rangle _{q_J}=0,}
\eqn\omegapmr{|\Omega _+\rangle _{q_J}\sim \prod _A \Psi _A^\dagger |\Omega _-\rangle _{q_J}, \qquad \hbox{and}\qquad  |\Omega _-\rangle _{q_J}\sim \prod _A \Psi _A |\Omega _+\rangle _{q_J}.} 
 Writing $``|0\rangle " _{q_J}$ as the naive (ignoring zero modes) groundstate for $q_J\neq 0$, 
\eqn\omegapmx{|\Omega _\pm \rangle _{q_J}\sim \left(\prod _{A}\Psi _{A}\right)^{\mp \half}``|0\rangle "_{q_J}.} 
 We identify the $X_\pm$ quanta with the top and bottom vortex states:
 \eqn\omegax{|\Omega _+ \rangle _{q_J=\pm 1} \sim X_{\pm }|0\rangle \qquad\hbox{and}\qquad |\Omega _-\rangle _{q_J=\pm 1}\sim X_{\mp}^\dagger |0\rangle,}
 with $|0\rangle$ the $q_J=0$ vacuum.       We verify that the vortex charges, computed from \omegapmx, are indeed compatible with \omegax\ and the $X_\pm$ charges in \onengen.  
 
 This matching was verified in \aspects\ for theories with $N$ matter fields $Q_i$, with all $n_i=1$. 
 The $N=1$ case is the classic $\CN =2$ susy Abelian Higgs model\foot{Though the Chern-Simons term must be included, $k\in \Z +\half$, reflecting the parity anomaly.}, and its 
vortices and zero modes have been studied in e.g.  \refs{\LeeEQ\LeeYC\LeePM\GoldhaberKN\MezincescuGB\KimUW\SchroersZY\CollieMX -\OikonomouTSA }.     Its $|q_J|=1$ vortex has one complex Fermion zero mode, $\Psi _1$ and \fermiquant\ leads to the BPS or anti-BPS doublet, $|\Omega _\pm \rangle _{q_J}$.  For $N>1$, there is a subtlety \aspects: the extra matter fields lead to $\sim 1/|z|$ non-normalizable (log-IR-divergent) Bose and Fermi zero modes, generalizing those found in \refs{\WardIJ\VachaspatiDZ\HindmarshYY-\LeeseFN}.  The $\rho _i$ terms in \qwind, allowed for matter with $Q_i^{vac}=0$, are examples; the interpretation of \aspects\ is that they are actually {\it vacuum} parameters. The matching \omegax\ requires that the non-normalizable Fermi zero modes nevertheless be included among the quantized $\Psi _A$ in \fermiquant\ and \omegapmx.

We here extend the analysis to theories with general matter charges $n_i$.   We find that, in the $q_J=\pm 1$ vortex background (for $\zeta >0$), the Fermion component of $Q_i$ leads to $|n_i|$ zero modes, $\Psi _{i, p=1\dots |n_i|}$,  with charges\foot{Here $U(1)_{\rm gauge}$ is Higgsed, so 
the $U(1)_{gauge}$  charges given here are screened by the $\ev{Q_i}$.} and spin given by (again, $k_c\equiv \half \sum _i n_i|n_i|$): 
\eqn\fermigen{
\vbox{\offinterlineskip\tabskip=0pt
\halign{\strut\vrule#
&~$#$~\hfil\vrule
&~$#$~\hfil\vrule
&~$#$~\hfil\vrule
&~$#$~\hfil\vrule
&~$#$~\hfil\vrule
&~$#$~\hfil\vrule
&~$#$~\hfil
&\vrule#
\cr
\noalign{\hrule}
& & U(1)_{\rm gauge} & U(1)_{\rm spin} & U(1)_j & U(1)_R &  U(1)_J \cr
\noalign{\hrule}
&  ``|0\rangle" _{q_J=\pm 1}  &\; \mp k &\; -\half k & \; 0 &\;0& \quad \pm 1   \cr
&  \Psi^{(q_J=\pm 1)}_{i, p=1\dots |n_i|}        &\; n_i &\; \pm {n_i\over |n_i|}(p-\half) & \;\delta _{ij} &\; -1& \quad 0   \cr
&\prod _{i, p}\Psi ^{(q_J=\pm 1)}_{i,p} &\;  2k_c &\;  \pm k_c&\;  |n_j|  &\; -\sum _i |n_i|  & \quad 0 \cr
&|\Omega _\pm \rangle _{q_J=1}&\;  \mp k_c -k&\;  \mp \half k_c-\half k &\;  \mp \half |n_j|  &\; \pm \half \sum _i |n_i|  & \quad 1 \cr
}\hrule}}
Quantizing the $\Psi _{i, p}$ gives a tower of $2^{\sum |n_i|}$ degenerate vortex states.  The 
top and bottom states $|\Omega _\pm \rangle _{q_J}$, as in \omegapm, have quantum numbers 
that follow from  \fermigen\ and \omegapmx; this gives the charges of $|\Omega _\pm \rangle _{q_J=1}$ in \fermigen.  These  $|\Omega _\pm \rangle _{q_J=1}$ charges indeed agree with those of $X_+$ and $X_-^\dagger$ in \onengen, fitting with the proposed operator / state map in \omegax. 

As we will see, the  $|q_J|=1$ Fermi zero modes in \fermigen\ have large $z$ behavior (from  \awind) $|\Psi _{i, p}|\sim |z|^{p-1-|n_i|}$, and the $p=|n_i|$ case is non-normalizable, for every matter field.  As in \aspects, we quantize all Fermi zero modes as in \fermiquant, including the non-normalizable ones, and interpret the non-normalizable Fermi zero modes as mapping between different Hilbert spaces.  But some additional discussion is required here, particularly for theories with $k=k_c=0$.  Then both $X_+$ and $X_-$ exist in the same theory, corresponding to the two Coulomb branches. Fitting with \omegax, both 
$|\Omega _+\rangle _{q_J=1}$ and $|\Omega _-\rangle _{q_J=1}$ in \fermigen\ have $U(1)_{\rm spin}$ zero, and can condense to give the $X_+$ or $X_-$ branches.  But $|\Omega _+\rangle _{q_J=1}$ and $|\Omega _-\rangle _{q_J=1}$ are are related via  non-normalizable Fermi zero modes.
 The BPS quanta created by $X_+$ and $X_-^\dagger$ evidently must reside in different Hilbert spaces, which seems puzzling.

Our (tentative) interpretation is that this reflects the fact that $X_+$ and $X_-$ label two disconnected branches of the moduli space of vacua, i.e. that $X_+X_-\sim 0$ in the chiral ring.  Quantum field theories typically do not have a Hilbert space of single-particle states, with a mapping between them via normalizable zero modes.   To the extent that it can happen for BPS states relies on the $x$-independence of the chiral ring OPE.  If a product of chiral operators is zero in the chiral ring, the associated BPS states can appear to reside in  different Hilbert spaces.    We discuss this further in sect. 5, e.g. for $N_f=1$ SQED, and its $W=MX_+X_-$ dual.   It would be good to have a more complete understanding.

The outline of the remaining sections is as follows.  Section 2 briefly reviews some of the basic points, and sets up our notation and conventions; a few more details are in an appendix. 
Section 3 broadly discusses the BPS vortices, and their zero modes, for the general $\CN =2$ susy, $U(1)_k$ charge $n_i$ matter theories. Section 4 discusses vortices and zero modes in general cases with a vev $\ev{Q_i}\propto \delta _{i,1}$, with $Q_1$ of charge $n_1=1$.   Section 5 considers theories with $N_\pm$ matter fields of charge $n_i=\pm 1$, e.g. $\CN =2$ SQED with $N_+=N_-=N_f$ flavors.  Section 6 discusses cases where $\ev{Q_i}\neq 0$ for matter with charge $n_i\neq 1$, where there can be an unbroken $\Z _{|n_i|}$ discrete gauge symmetry, i.e. an orbifold. 

One could generalize to non-Abelian gauge theories; it will not be considered here.

\newsec{A few preliminaries (see also the appendix)}

\subsec{Lagrangian and effective Chern-Simons terms}
The $U(1)_k$ gauge theory, with matter fields $Q_i$ of charges $n_i$, has classical Lagrangian
\eqn\lclass{\CL _{cl}\supset \int d^4\theta \left(-{1\over e^2}\Sigma ^2-{k\over 4\pi} \Sigma V-{\zeta \over 2\pi} V+\sum _j Q_j^\dagger e^{2n_jV+2im_j\theta \bar\theta }Q_j\right).}
We will set the real masses $m_i=0$, and take $W_{tree}=0$.  Dirac-quantization for monopole operators implies that the Chern-Simons coefficient $k$ is quantized as
\eqn\consk{k + \half \sum_i n_i^2 \in \Z; \qquad\hbox{equivalently,} \qquad  k + \half \sum_i n_i \in \Z~~.}

The supersymmetric vacua have expectation values of the Coulomb modulus $\sigma = \Sigma |$, or the matter fields $Q _i=Q_i|$, subject to the conditions $D=0$ and $m_j(\sigma )Q_j=0$, where 
\eqn\dis{D=-e^2(\sum _i n_i |Q_i|^2-{\zeta _{eff}\over 2\pi}-{k _{eff}\over 2\pi}\sigma) ,}
and $m_i(\sigma)\equiv m_i +n_i\sigma$.  The effective FI parameter $\zeta _{eff}$, and Chern-Simons coefficient $k_{eff}$ in \dis\ are shifted by integrating out massive matter, with $\zeta _{eff}=\zeta$ for $m_i=0$ and 
\eqn\keff{k_{eff} (\sigma) =k + \half \sum_i n_i^2 \sign(m_i(\sigma)) \in \Z.}
``Higgs"  susy vacua have $(Q _i\neq 0, \ \sigma =0$), while ``Coulomb" vacua have $(Q_i=0, \  \sigma \neq 0$) and $k_{eff}=\zeta _{eff}=0$.  The asymptotic values  of $k_{eff}$ for $\sigma \to \pm \infty$ are 
\eqn\kinfinity{k_{eff}(\sigma =\pm \infty)=k\pm k_c, \qquad k_c\equiv \half \sum _i n_i |n_i|.}
So the $\sigma \to \pm \infty$  asymptotic regions of the Coulomb branch only exist if  \kinfinity\ vanishes, i.e. if $k=\mp k_c$, respectively.  For non-zero $k_{eff}$ and $\zeta _{eff}$, there are also isolated ``topological vacua," with $Q _i=0$ and $\sigma =-\zeta _{eff}/k_{eff}$; those vacua will not enter in our discussion. 

\subsec{Chern-Simons contribution to Gauss' law, and charges and spin from $q_J$}
The Chern-Simons term affects Gauss' law (the $A_0$ EOM), as 
 \eqn\gaussis{-{1\over e ^2}\partial _i F_{0i}=\rho_{matter}-{k\over 2\pi}F_{12},}
 with $\rho _{matter}\equiv {\delta \CL _{matter}\over \delta A^0}$ the matter contribution to the electric charge density (see the appendix for our sign conventions).  The ``$k$" in \gaussis\ is the classical value when we consider the theory at $\sigma =0$, where the matter Fermions are massless and kept in the low-energy theory.  On the other hand, for $\sigma \neq 0$, the matter Fermions are massive and can be integrated out, and then we should replace $k$ in \gaussis\ with $k_{eff}$, as in \keff.   
 
The Chern-Simons contribution in \gaussis\ implies that operators or states with $q_J\neq 0$ acquire an associated electric charge, and a related 
contribution to their spin \refs{\WilczekDU\JackiwAW-\JackiwPR}
\eqn\chargespin{q_{elec}=-kq_J, \qquad \Delta s = -\half k q_J^2; }
with $k\to k_{eff}$ in \chargespin\ if the Fermions are massive and integrated out.
For vortices, if $k\neq 0$, the last term  in \gaussis\ leads to $A_0\neq 0$, which complicates 
the equations of motion.

The gauge and global charges of the $X_\pm$ operators in \onengen\ follow from \chargespin, and its analogs for mixed gauge-flavor Chern-Simons coefficients.  Since $X_\pm$ extend to $\sigma = \pm \infty$, we replace $k\to k_{eff}(\sigma = \pm \infty)=k\pm k_c$ \kinfinity, and use 
 $q_J\to \pm 1$ in \chargespin\ to obtain the $U(1)_{\rm gauge}$ charges of 
 $X_\pm$ in \onengen.  The $\sigma \to \pm \infty$ Coulomb branch only exists if $k_{eff}=0$, which is the condition for $X_\pm$ to be a gauge invariant, scalar operator: 
\eqn\coulombexists{\hbox{If $k=\mp k_c$,  then the $X_\pm$ Coulomb branch exists.}}
The $U(1)_i$ and $U(1)_R$ global charges of $X_\pm$ in \onengen\ likewise follow immediately from the one-loop  induced, mixed Chern-Simons terms between the gauge field and  background gauge fields coupled to the global currents \refs{\AharonyBX, \ClossetVG, \ClossetVP, \aspects}.  Integrating out the matter Fermion components of $Q_i$ in \onengen,  of mass $m_i(\sigma)=n_i\sigma$, gives mixed  CS terms $k_{eff}^{gauge, U(1)_j}=\half n_j \sign (n_j\sigma)$ and $k_{eff}^{gauge, U(1)_R}=-\half \sum _i n_i \sign (n_i \sigma)$. Taking $\sigma \to \pm \infty$ for $q_J=\pm 1$, the analog \chargespin\ for the global charges then gives the corresponding charges in \onengen. 

\subsec{BPS and anti-BPS particles}

Particle states can be labelled by their $U(1)_{spin}$, $s$, and it is convenient to convert the spinors to a rotational spin-diagonal basis  ($s=1$ for $z=x^1+ix^2$ and $\partial _{\bar z}=\half (\partial _{x^1}+i\partial _{x^2})$).  For the supercharges, we define (fixing a minor notational issue vs \aspects)
\eqn\Qpmd{Q_\pm\equiv \half(Q_1\mp i Q_2), \qquad \bar Q_\pm\equiv \overline{Q_\mp}=\half(\bar Q_1\mp i \bar Q_2),}
so $Q_\pm$ and $\bar Q_\pm$ have spin $s=\pm\half$.  In terms of these, the ${\cal N}=2$ algebra is
\eqn\zenopm{\{Q_\pm, \bar Q_\pm\} =\mp i (P_1 \pm i P_2) , \qquad
\{Q_\pm, \bar Q_\mp\} =  P^0 \pm Z.
}

A BPS particle, with $m=Z$, has
\eqn\bpsa{Z>0: \qquad  Q_-|BPS\rangle = \bar Q_+|BPS\rangle =0,}
and the remaining two supercharges make a two-dimensional representation
\eqn\bpsp{Z>0: \qquad |BPS\rangle = \pmatrix{|a\rangle \cr |b\rangle }, \qquad \bar Q_-|a\rangle =0, \qquad |b\rangle = Q_+|a\rangle.}
Likewise, an anti-BPS particle has $m=-Z>0$,  and is annihilated by $Q_+$ and $\bar Q_-$.  Every BPS state has a CPT conjugate anti-BPS state, with opposite global charges and $Z$, but with the same $U(1)_{spin}$ spin $s$.  The R-charges and spins of these states are \aspects 
\eqn\spectrum{
\vbox{\offinterlineskip\tabskip=0pt
\halign{\strut\vrule#
&~$#$~\hfil\vrule
&~$#$~\hfil\vrule
&~$#$~\hfil\vrule
&~$#$~\hfil\vrule
&~$#$~\hfil\vrule
&~$#$~\hfil
&\vrule#
\cr
\noalign{\hrule}
&  & U(1)_R &  U(1)_{spin} & \; Z\cr
\noalign{\hrule}
&  |a\rangle      &\;r& \quad s  & \;>0\cr
&|b\rangle  &\; r-1  & \; s+\half & \;>0 \cr
&  |\bar a\rangle      &\;-r& \quad s & \;<0  \cr
&|\bar b\rangle  &\; -r+1  & \; s+\half & \;<0\cr
}\hrule}}

\newsec{BPS and anti-BPS vortices}

The central term of the supersymmetry algebra (setting real masses $m_i=0$) is
\eqn\zis{Z=\zeta q_J.}
For $Z>0$, the vortex can be BPS, annihilated by $Q_-$ and $\bar Q_+$ \bpsa.  For $Z<0$, the vortex is anti-BPS, annihilated by $\bar Q_-$ and $Q_+$.  The condition that these supercharges annihilate the background implies the BPS equations for a  {\it static} (all $\partial _t\to 0$) vortex with $Z>0$ (resp, a $Z<0$ anti-BPS vortex) are (with $D_{z}^{(n_j)}\equiv \half (D_1-i D_2)^{(n_j)}\equiv \partial _{z}+i n_j A_{z}$)
\eqn\bpst{\sigma = \pm A_0,}
\eqn\finaleq{F_{12}=\pm D,} 
\eqn\bpsphi{D^{(n_j)}_zQ_j=0, \qquad \hbox{resp}\qquad D^{(n_j)}_{\bar z}Q_j=0,}
with $D$ given by \dis.  One must also impose Gauss' law \gaussis.   In our conventions, the chiral superfields, $Q_i$,  of a $Z>0$ BPS vortex are {\it anti}\foot{This (unfortunately) is due to following \WessCP 's sign convention for $A_\mu$; see the appendix. Fitting with 
\zenopm\ and \bpsa, $\{Q_-, \bar Q_-\} =2iP_z$ annihilates the BPS chiral field configuration, since chiral fields  are annihilated by $\bar Q_\pm$, and BPS configurations by $Q_-$ and $\bar Q_+$.  Compared to e.g. \refs{\LeeEQ, \LeeYC, \LeePM}, $(A_\mu, \sigma , \lambda _\alpha, \bar \lambda _{\alpha})^{here}=-e(A_\mu , N, \lambda _\alpha , \chi _\alpha)^{there}$, $q_J^{here}=-n_{there}$.}-holomorphic (resp holomorphic for a $Z<0$ anti-BPS vortex).    We will here be particularly interested in the zero modes.

The  vortex's Fermi zero modes are the static $\partial _t\to 0$ solutions of the Fermion equations of motion, from \lclass\ with $m_i=0$, in the background of the static vortex's Bosonic fields:  
\eqn\fermizmi{\pmatrix{i{k\over 4\pi} & 2e^{-2}\partial _{\bar z}\cr -2 e^{-2}\partial_z & i{k\over 4\pi}}\pmatrix{\bar \lambda _{\uparrow}\cr \bar \lambda _{\downarrow}}-\sqrt{2}\sum _j n_j Q _j^*\pmatrix{\psi _{j\uparrow}\cr \psi _{j\downarrow}}=0,}
\eqn\mattzm{\pmatrix{i n_j (A_0-\sigma)& 2D^{(n_j)}_{\bar z}\cr -2D^{(n_j)}_z & -in_j (A_0+\sigma)}\pmatrix{\psi _{j\uparrow}\cr \psi _{j\downarrow}}-\sqrt{2} n_j Q _j \pmatrix{\bar \lambda _{\uparrow}\cr \bar\lambda _{\downarrow}}=0,}
where $\psi _{i\uparrow, \downarrow}$ and $\bar \lambda _{\uparrow, \downarrow}$ have spin $\pm \half$, and $U(1)_R$ charge $-1$.  As we discuss in section 4, the number of solutions of \fermizmi\ and \mattzm, and their quantum numbers, are as in \fermigen: each matter field contributes $|n_i|$ Fermi zero modes, with spin correlated to the sign of $n_i$.   

\subsec{Review of the minimal matter example: a single matter field $Q_1$ of charge $n_1=1$}

This is the basic $\CN =2$ Abelian Higgs model, and its BPS vortices have been discussed e.g. in \refs{\JackiwAW, \JackiwPR, \LeeEQ, \LeePM\GoldhaberKN -\MezincescuGB}.  We here review the discussion from \aspects.  By \consk, here $k\in \Z +\half$, and the theory has $\Tr (-1)^F=|k|+\half$ vacua \aspects; we here discuss the BPS vortices of the theory in the Higgs\foot{For $|k| >\half$, one could consider vortices in the other vacua, with $\ev{Q_1}=0$ and $\ev{\sigma}\neq 0$, and domain walls between the vacua, as in \JackiwPR, but we will not consider such configurations here.}  vacuum of the theory with FI parameter $\zeta>0$, i.e. $\ev{Q _1}=\sqrt{\zeta /2\pi}$.

The solution  $A^{vortex}_\mu (z, \bar z)$, $Q^{vortex}_1(z, \bar z)$, $\sigma ^{vortex}(z, \bar z)$ of the BPS field equations \bpst, \finaleq, \bpsphi, is not analytically known, nor is it needed: knowing its existence and number of zero modes suffices. The vortex with $U(1)_J$ charge $q_J$ has $|q_J|$ complex Bosonic zero modes, and $|q_J|$ spin $+\half$ Fermionic zero modes.   
The $q_J=1$ vortex has one complex zero mode $z_1$, the translational invariance zero mode of the BPS vortex core location, and one 
complex spin $\half$ Fermionic zero mode \refs{\LeePM, \GoldhaberKN}, $\Psi _1$, a combination of the photino and the matter fermion that solves \fermizmi\ and \mattzm.  The Bosonic field configuration is  annihilated by $Q_-$ and $\bar Q_+$ \bpsa, while the other two supercharges give the Fermi zero mode, $\Psi _1\sim Q_+$,  and complex conjugate $\Psi _1^\dagger \sim \bar Q_-$, i.e. the photino  and matter Fermi 
field configuration of $\Psi _1$ follows from acting with $Q_+$ on  $F^{vortex}_{\mu \nu}(z, \bar z)$ and $Q^{vortex}_1(z, \bar z)$.   

Quantizing the $q_J=1$ vortex $\Psi _1$ Fermi zero mode, $\{\Psi _1, \Psi _1^\dagger\}=1$ (so $\Psi _1\to Q_+/\sqrt{2E}$) yields a BPS doublet  \bpsp; 
adding the $q_J=-1$, anti-BPS, CPT conjugate states gives one copy of the spectrum \spectrum. The $U(1)_R$ and $U(1)_{\rm spin}$ quantum numbers there are found as in \omegapmx\ from those of $\Psi _1$, $|\Omega _\pm \rangle _{q_J=1}\sim \Psi _1^{\mp \half} ``|0\rangle "_{q_J=1}$ with $``|0\rangle "_{q_J=1}$ assigned spin $-\half k$ as in \chargespin.  This gives $r=\half$ and  $s=-\half (k+\half )$ \aspects, as in \fermigen\ with $k_c=\half \sum _i n_i |n_i|=\half$.  The $k=\mp \half$ theory is dual to a theory of a free chiral superfield, $X_\pm$ \DimofteJU.  The FI parameter $\zeta$ maps to a real mass $m_X$ in the dual.  BPS vortices map to $X$-particle states. 

\subsec{Cases with multiple matter fields $Q_i$: the (anti)-BPS equations for the Bosonic fields}

By \bpsphi, the vortex gauge field configuration is completely determined by that of {\it any} non-zero matter field $Q_i$:
\eqn\asolved{A_z= {i\over n_i}\partial _z\ln Q_i, \qquad\hbox{resp}\qquad A_{\bar z}={i\over n_i}\partial _{\bar z}\ln Q_i, \qquad \hbox{for any}\quad Q_i\neq 0.}
The condition that the gauge field \asolved\ be smooth, with winding number $q_J$ \awind, implies \TaubesTM\ that  a charge $n_i=1$ matter field has $Q_i(z)$ with $|q_J|$ zeros, at the vortex core locations, $z=z_{i=1\dots |q_J|}$.  For $|q_J|=1$, a charge $n_i$ matter field with 
$Q_i^{vac}\neq 0$ can have an 
order $|n_i|$ zero at the location $z_1$ of the BPS (resp. anti-BPS) vortex core 
\eqn\qszero{Q_i^{vac}\neq 0:\qquad Q_i = (\bar z-\bar z_1)^{|n_i|}  f_i, \qquad \hbox{resp}\qquad Q_i=(z-z_1)^{|n_i|} f_i,}
with $f_i\equiv f_i(z, \bar z)$ non-vanishing.   Turning on Bosonic zero modes can resolve the zeros in \qszero\ or, with multiple matter fields, eliminate the zeros, as in the examples of \refs{\VachaspatiDZ \HindmarshYY-\LeeseFN}.

Using \asolved, the BPS equations  \bpsphi\  can be rewritten in terms of ordinary derivatives  and $U(1)_{gauge}$ neutral ratios of fields, where we divide by any $Q_i$ with $Q_i^{vac}\neq 0$: 
\eqn\bpsphiw{\hbox{(BPS):}\ \partial _z \left({Q_j \over Q_i ^{n_j/n_i}}\right)=0, \qquad\hbox{resp}\qquad \hbox{(anti-BPS):}\ \partial _{\bar z} \left({Q_j \over Q_i ^{n_j/n_i}}\right)=0.}

\subsec{Vanishing theorem and its consequences}

The non-zero solutions of \bpsphi\ are restricted by a vanishing theorem:
``{\it a line bundle of negative degree cannot have a non-zero holomorphic section}"; see e.g. \WittenYC\ for a nice discussion in the similar context of 2d instantons.  With our conventions, this implies
\eqn\vanishing{\eqalign{\hbox{BPS}:\quad Q_i&=0 \qquad\hbox{unless}\qquad \sign (n_i)=\sign (q_J)\cr \hbox{anti-BPS}:\quad Q_i&=0 \qquad\hbox{unless}\qquad \sign (n_i)=-\sign (q_J).}}
This can be seen from the identity (writing $x^\mu = (t, \vec x)$ and $D^{(n_j)}_\mu \equiv  (D^{(n_j)}_0, \vec D^{(n_j)})$)
\eqn\identity{\int d^2\vec x  |\vec D ^{(n_j)}Q _j |^2= \int d^2 \vec x \left(|2D^{(n_j)}_{z, \bar z} Q _j |^2\pm n_j |Q _j|^2 F_{12}\right);}
with $[D^{(n_j)}_{\bar z}, D^{(n_j)}_z]=\half n_j F_{12}$.  Since the LHS of \identity\ is non-negative, equations \bpsphi\ have a $Q_j\neq 0$ solution only if the second term on the RHS of \identity\ has the correct sign. By \zis, the $q_J\neq 0$  BPS vacua have $\sign (q_J)=\sign (\zeta)$ and the anti-BPS vacua have $\sign (q_J)=-\sign (\zeta)$.   So \vanishing\ implies, for both BPS and anti-BPS configurations
\eqn\vanishingii{Q_i=0 \qquad\hbox{if }\qquad \sign (n_i)=-\sign (\zeta).}
 
 An  immediate corollary is that there are only BPS vortices in Higgs vacua where $Q_i^{vac}$ satisfy \vanishingii, i.e. we solve $D=0$ \dis\ with $Q_i^{vac}\neq 0$ only for matter with $\sign (n_i)=\sign (\zeta)$.  So, in theories with matter fields with $n_i$ of both signs, all gauge-invariant products, i.e. the Higgs branch moduli, must be set to zero, e.g.  the  meson fields $M_{i\tilde j}= Q_i\tilde Q_{\tilde j}=0$ in a theory with vector-like matter.   
 As discussed in \AharonyBX, the fact that BPS vortices require $M_{i\tilde j}=0$ can have a simple dual perspective, e.g. for $N_f=1$ SQED  it is clear from the  
$W=MX_+X_-$ dual  that the $X_\pm$ quanta are only BPS for $M=0$.    See \refs{\PeninSI, \AchucarroII}\ for other, dynamical arguments leading to the same conclusion.

\subsec{Bosonic zero modes of $|q_J|=1$ BPS vortices with multiple matter fields}

Each matter field with $\sign (n_i)=\sign (\zeta)$ has $|n_i|$ complex Bosonic zero modes, one of which is the vortex core location, $z_1$ in \qszero.  Since matter fields with $\sign (n_i)=-\sign (\zeta)$ are set to zero \vanishingii, they do not yield Bosonic zero modes.    Consider \bpsphiw, taking say $Q_1$ and $Q_j$ to have  $\sign (n_1)=\sign (n_j)=\sign (\zeta)$, and suppose that $Q_1^{vac}\neq 0$ and $Q_j^{vac}=0$.  The general solution of \bpsphiw\ for a $q_J=1$ BPS (or $q_J=-1$ anti-BPS) vortex is then
\eqn\ratiosare{{Q_j(z, \bar z)\over Q_1(z, \bar z)^{n_j/n_1}}={\bar P_j (\bar z)\over (\bar z-\bar z_1)^{|n_j|}}, \qquad \hbox{resp}\qquad {Q_j(z, \bar z)\over Q_1(z, \bar z)^{n_j/n_1}}={P_j (z)\over (z-z_1)^{|n_j|}},}
where the denominators are determined by the $z\to z_1$ vanishing degree of $Q_1$ in \qszero, (which is the only singularity of the ratio) and the numerators by (anti) holomorphy and the condition that the ratio approaches the vacuum value, i.e. zero, for $|z|\to \infty$:
\eqn\pare{\bar P_j(\bar z)\equiv \sum _{p=1}^{|n_j|} \bar c_{j, p} \bar z ^{p-1}, \qquad\hbox{resp}\qquad P_j(z)\equiv \sum _{p=1}^{|n_j|} c_{j, p} z ^{p-1}.}
The $|n_j|$ coefficients $\bar c_{j, p}$ (or $c_{j, p}$) in \pare\ are the Bosonic zero modes for  matter field $Q_j$ with $Q_j^{vac}=0$ in a BPS (or anti-BPS) $q_J=1$ vortex. Matter field(s) $Q_i$ with $Q_i^{vac}\neq 0$ also yield $|n_i|$ Bosonic zero modes, one of which is the translational zero mode $z_1$.

\subsec{Normalizable vs non-normalizable zero modes}

The Bosonic or Fermionic zero modes of the static vortex are replaced with dynamical variables on the vortex worldline theory, {\it if} the associated induced kinetic term is normalizable.  Non-normalizable zero modes, on the other hand, are frozen parameters.   For example, the translational zero mode of a $|q_J|=1$ vortex is quantized as $z_1\to z_1(t)$, which is normalizable, with finite induced kinetic term $\int d^2 z{\cal L}\to \half m_{BPS}|\dot z_1|^2$.  Considering the $c_{j, p}$ or $\bar c_{j, p}$ term in \ratiosare\ for large $|z|$  gives $|Q_j|\sim |c_{j, p}| |z|^{p-1-|n_j|}$, so the induced coefficient of a $|\dot c_{j, p}|^2$ term involves $\sim \int d^2 z |z|^{2(p-1-|n_j|)}$, i.e. $c_{j, p}$ and $\bar c_{j, p}$ are normalizable  for $1\leq p<|n_j|$ (requiring $|n_j|>1$) and log-IR-divergent non-normalizable for $p=|n_j|$.  

The non-normalizable $\rho _j \equiv \bar c_{j, p=|n_j|}$ or $\rho _j \equiv c_{j, p=|n_j|}$ zero modes in \ratiosare\ generalize the non-normalizable zero modes of ``semi-local vortices" \refs{\WardIJ \VachaspatiDZ \HindmarshYY- \LeeseFN}.  As found there, turning on $\rho _i\neq 0$ dramatically changes the character of the vortex solution, removing the zero in \qszero\ at the vortex core, and changing the flux $F_{12}$ in \finaleq\ from having the usual $\sim e^{-c m_\gamma |z|}$ exponential falloff for large $|z|$ (with $m_\gamma$ the Higgsed photon mass) into a diffuse, power-law falloff.    In our general $n_i$ case, each matter field with $\sign (n_i)=\sign (\zeta)$ and $Q_i^{vac}=0$ yields one-such non-normalizable $\rho _i$ bosonic zero mode.  If $|n_j|>1$, there are also $|n_j|-1$ additional normalizable, and hence dynamical, zero modes $\bar c_{j, p<|n_j|}$ or $c_{j, p<|n_j|}$.   

The bosonic non-normalizable zero modes, $\rho _i$, are interpreted, as in \aspects, as 
superselection parameters already of the $q_J=0$ {\it vacuum}, even before adding the vortex:
taking  $Q_i \sim \rho _i /|z|$ for large $|z|$ has finite energy, with $\rho _i$ non-normalizable, so unchanging in time. 
Likewise, Fermi zero modes are either normalizable, if $\Psi _A <\CO(1/|z|)$ for large $|z|$, or non-normalizable if $\Psi _A =\CO(1/|z|)$.  As in \aspects, we quantize {\it all} the Fermion zero modes as in \fermiquant, including the non-normalizable ones.  The 
tower of $2^{N_z}$ states discussed around \fermiquant\ includes states in different Hilbert spaces, if related by a non-normalizable Fermi zero mode.  The charges of the states, and in particular the states $|\Omega _\pm\rangle _{q_J}$ at the top and bottom of the tower, are affected by all the Fermi zero modes, with the product in \omegapmx\ including all normalizable and also non-normalizable Fermi zero modes.

\newsec{Fermi zero modes of BPS vortices for somewhat general cases.}

We will consider $|q_J|=1$ BPS and anti-BPS vortices, taking $\zeta /n_1 >0$, in the vacuum with 
$\ev{\sigma}=0$ and non-zero expectation value for only $Q_1$:
\eqn\qivac{ \qquad Q_i^{vac}=\sqrt{\zeta \over 2\pi n_1}\delta _{i,1}.}
For the rest of this section, we assume that $n_1=1$, though we allow for general charges $n_j$ for the other $Q_{j>1}$ matter fields in \qivac. We will discuss the $n_1\neq 1$ case in sect. 6.  

  Each $Q_i$ matter field with $n_i>0$ has $n_i$ Bosonic zero modes, while $Q_i$ with $n_i<0$ have none. 
The $Q_1$ Bosonic zero mode is the normalizable, translational zero mode, $z_1$.  For the matter fields $Q_{j\neq 1}$, with $n_j>0$,  the Bosonic zero modes are the $\bar c_{j, p}$ or $c_{j, p}$ in \pare, with $p=|n_j|$ non-normalizable.   Non-zero time derivatives of the normalizable  $c_{j, p}$ and  $\bar c_{j, p}$ can contribute to the  vortex's energy, momentum, and spin angular momentum.

We now consider the Fermi zero modes of the $q_J=1$ BPS vortex or $q_J=-1$ anti-BPS vortex in the vacuum \qivac.  Since the counting and quantum numbers of Fermi zero modes cannot depend on continuous variables, we can simplify things by setting all Bosonic zero modes to zero, in which case
\eqn\qivortex{Q_i^{vortex}(z, \bar z)=Q_1^{vortex}(z, \bar z)\delta _{i,1}.}
Here $Q_1^{vortex}$ coincides with that of $U(1)_k$ with only the matter field $Q_1$; the $Q_{i\neq 1}$ matter fields do not affect the solution.   Likewise, 
the Fermi zero mode equations \fermizmi\ and \mattzm\ involving $\bar \lambda _\pm$ and $\psi _{1\pm}$ decouple from those for the $Q_{j>1}$ matter Fermions.  The solution for the zero modes from $\bar \lambda _{\uparrow, \downarrow}$ and $\psi _{1\uparrow, \downarrow}$ is the same as that of the minimal matter theory reviewed  in sect. 3.1: for $|q_J|=1$ it gives one 
Fermion zero mode, $\Psi _1$, and conjugate $\Psi _1^\dagger$, corresponding to the  non-trivial supercharges $Q_+$ and $\bar Q_-$ in \bpsp.   

Now consider the decoupled equations \mattzm\ for the $Q_{j>1}$ matter Fermi zero modes:
\eqn\mattzmx{\eqalign{\hbox{BPS} (j\neq 1):\qquad D^{(n_j)}_{\bar z}\psi _{j\downarrow}&=0, \qquad D_z^{(n_j)} \psi _{j\uparrow}=-in_j A_0\psi _{j\downarrow}; \cr 
\hbox{anti-BPS} (j\neq 1):\qquad D^{(n_j)}_{z}\psi _{j\uparrow}&=0, \qquad D_{\bar z}^{(n_j)} \psi _{j\downarrow}=-in_j A_0\psi _{j\uparrow}.}} 
  For $k=0$, it is possible to set $\sigma = A_0=0$, and we obtain the simpler version
\eqn\baby{\hbox{($j\neq 1$ simple version)}\qquad D^{(n_j)}_z \psi _{j, \uparrow}=0, \qquad\hbox{and}\qquad D^{(n_j)}_{\bar z} \psi _{j, \downarrow}=0.}
If $k\neq 0$, Gauss' law \gaussis\ implies that $A_0=\pm \sigma$ is a complicated function.  Fortunately, for any value of $k$, \mattzmx\ and the simpler version \baby\ have the same number of zero mode solutions, with the same spins.  Indeed, using \identity\ and \vanishing, it follows that 
\eqn\vanishingf{\cases{D^{(n_j)}_{\bar z}\psi _{j,\downarrow}=0\to \psi _{j\downarrow}=0& if  $n_j q_J>0$\cr D^{(n_j)}_{z} \psi _{j, \uparrow}=0 \to \psi _{j, \uparrow}=0& if $n_j q_J<0$}.}

Consider the case of a $q_J=+1$ BPS vortex; the anti-BPS case is analogous.  For matter with $n_j>0$,  \vanishingf\ gives $\psi _{j\downarrow}=0$ and \mattzmx\ reduces to \baby.  For $ n_j<0$ matter, $\psi _{j\downarrow}$ is non-trivial and satisfies the same equation in  \mattzmx\ and \baby.  The difference between the $\psi _{j\uparrow}$ equations in \mattzmx\ and \baby\  for $n_j<0$ is immaterial in terms of counting solutions: the solution for $\psi _{j\uparrow}$ in either equation is uniquely determined, as $D_z^{(n_i)}$ has trivial kernel for $n_i<0$ \vanishingf.  So we can always count Fermi zero modes via  \baby. 

Using \asolved\ to eliminate the gauge field in favor of $Q_1$, the equations \baby\ become 
\eqn\fratio{\hbox{if}\quad n_j>0: \quad 
\partial _z \left({\psi _{j, \uparrow}\over Q_1^{n_j}}\right)=0;
\qquad  \hbox{if}\quad n_j<0: 
\quad \partial _{\bar z}\left({\psi _{j, \downarrow}\over (Q_1^\dagger)^{|n_j|}}\right)=0.
}  
Since, for $q_J=1$, $Q_1$ has a degree one zero at $z_1$, this gives (similar to \ratiosare)
\eqn\fplusis{n_j>0 \quad (q_J=1):\qquad \psi _{j, \uparrow}={Q_1^{n_j}\over (\bar z-\bar z_0)^{n_j}}\sum _{p=1}^{n_j} \bar u_{j, p} \bar z^{p-1},}
with the $n_j$ coefficients, $\bar u_{j, p=1, \dots n_j}$, Fermionic zero modes of spin $p-\half $.  Likewise,
\eqn\fminusis{n_j<0\quad (q_J=1):\qquad \psi _{j, \downarrow}={(Q_1^\dagger)^{|n_j|}\over ( z-z_0)^{|n_j|}}\sum _{p=1}^{|n_j|} d_{j, p} z^{p-1},}
with the $|n_j|$ coefficients, $d_{j, p}$, Fermionic zero modes of spin $-(p-\half )$.  As in the bosonic case, for either \fplusis\ or \fminusis, the $p=|n_j|$ Fermi zero mode is non-normalizable.  
The spins of $\bar u_{j, p}$ and $d_{j, p}$ follow from constructing the angular momentum generator, much as in  \KimYZ, assigning spin $+1$ to $z$, and 
spin $+\half$ to $\psi _{j, \uparrow}$ in \fplusis.  By \qwind, $Q_1^{n_j}/(\bar z-\bar z_0)^{n_j}$ is $\theta$ independent for large $|z|$, so we assign spin $+\half$ to each term $\bar u_{j,p}\bar z^{p-1}$ in \fplusis, and, likewise,  spin $-\half$ to all $d_{j, p}z^{p-1}$ in \fminusis.  So $\bar u_{j,p}$ has spin $p-\half$ and $d_{j, p}$ has spin $-(p-\half)$. In sum, the $q_J=1$ vortex has the $\Psi _{n_j, p}^{(q_J=1)}$ in \fermigen: $|n_j|$ Fermion zero modes, of spins $\sign (n_j)(p-\half)$, for $p=1\dots |n_j|$.  The $q_J=-1$ vortex is similar.   The 
other quantum numbers likewise follow from those of $\psi _{j, \uparrow, \downarrow}$, and are as given in \fermigen.   We assign $U(1)_{\rm gauge}$ charges in \fermigen, even though $U(1)_{\rm gauge}$ is spontaneously broken (screened) by \qivac.  

The Bose and Fermi zero modes form supermultiplets of a 1d worldline theory with two unbroken supercharges (see e.g. \refs{\CollieMX, \SamolsNE, \KimQMA}\ for some examples), as in the  1d reduction of a 2d $\CN =(2,0)$ worldsheet theory of  BPS vortex strings in 4d $\CN =1$ theories  \refs{\EdalatiVK, \TongQD, \ShifmanCE}.  The zero modes of a matter field $Q_i$ are in $|n_i|$ different $\CN =(2,0)$ chiral multiplets (i.e. a complex Boson and a complex Fermion) if $\sign (n_i)=\sign (\zeta)$, or $|n_i|$  $\CN =(2,0)$ chiral Fermi multiplets (i.e. a complex Fermion and an auxiliary field) if $\sign (n_i)=-\sign (\zeta)$.   

All the Fermi zero modes are quantized, as in \fermiquant\ and \omegapm, giving $2^{\sum |n_i|}$ states.  The $\Psi ^{(q_J=1)}_1\sim Q_+ Q^{vortex}_{i=1}$ zero mode should be regarded as $Q_+$, i.e. neutral under $U(1)_{gauge}$ and the non-R-symmetry global symmetries; quantizing this zero mode 
yields BPS doublets \spectrum.   Including all zero modes yields $2^{\sum |n_i|-1}$ BPS doublets. 

Consider a theory with vector-like, charge-conjugation  symmetric matter content, with pairs $Q_i$ and $\tilde Q_i$, of charges $\pm n_i$.  Then $k_c=\half \sum _i n_i |n_i|=0$ in \kinfinity, and the $k=0$ theory with $\zeta =0$ has asymptotic Coulomb branches $X_\pm$.  The theory respects $P$  and $T$ if $k=0$, and it respects $C$ if $\zeta =0$.   For every Fermi zero mode $\Psi _{n_j, p}$, there is a Fermi zero mode $\tilde \Psi _{-n_j, p}$ of opposite spin, so the $\prod _A \Psi _A$ appearing in \omegapmx\ has spin $s=0$, and the top and bottom states $|\Omega _\pm \rangle _{q_J=1}$ have $s=-\half k$, so spin $0$ for $k=0$,   This fits with  \omegax: these states map to the quanta of $X_\pm$,   $|\Omega _+\rangle _{q_J=\pm 1}\sim X_\pm |0\rangle$ and $|\Omega _-\rangle _{q_J=\pm 1}\sim X^{\dagger}_{\mp }|0\rangle$, with $X_\pm$ a gauge invariant operator for $k=0$.

\newsec{Examples: theories with $N_\pm $ matter fields of charge $n_i=\pm 1$}

We denote the matter  as $Q_{i=1\dots N_+}$, with $n_i=+1$, and $\tilde Q_{\tilde i =1\dots N_-}$, with $n_{\tilde i}=-1$.  The $U(1)_j$ global symmetries in \onengen\ enhance to $SU(N_+)\times SU(N_-)\times U(1)_A$, where the $U(1)_A$ charge is $+1$ for all $Q_i$ and $\tilde Q_{\tilde i}$.  We take $N_+>0$, and $\zeta >0$, and then \qivac\ is the general vacuum with BPS vortices; it spontaneously breaks $SU(N_+)\to SU(N_+-1)\times U(1)$, so for $N_+>1$ the vacua contain the NG bosons $\cong CP^{N_+-1}$.  For $N_+N_-\neq 0$, the vacua also include non-compact directions, given by the mesons $M_{i\tilde j}=Q_i\tilde Q_{\tilde j}$, with $M_{i\tilde j}$ of rank 1, but as in \vanishingii\ BPS or anti-BPS vortices require 
 $\tilde Q_{\tilde i}=0$, so $M_{i\tilde j}=0$.   The Chern-Simons quantization condition \consk\ gives $k+\half \Delta N\in \Z$, with $\Delta N\equiv N_+-N_-$; also, $k_c=\half \Delta N$. 

The cases $(N_+, N_-)=(N, 0)$ were discussed in \aspects.  The minimal matter case, $N=1$, was reviewed in sect. 3.1.  The vortices of the $N>1$ case is 
 the $\CN =2$ version of the  ``semi-local" vortices of \refs{\WardIJ \VachaspatiDZ- \HindmarshYY}, allowing also for Chern-Simons terms.   Our present discussion in this section also includes cases with both $N_+N_-\neq 0$; we did not find much discussion of vortices in such theories in the literature, aside from some brief comments in \refs{\KimUW, \SchroersZY}. 

For general $(N_+, N_-)$, a $q_J=1$ BPS vortex has $N_+$ complex bosonic zero modes.  One is the normalizable, translational zero mode, $z_1$, corresponding to the vortex core location.  The remaining $N_+-1$ bosonic zero modes are the non-normalizable $\bar \rho _i$ parameters in
\eqn\bozos{ {Q_{i\neq 1}\over Q_1}={\bar \rho _i\over \bar z}.}
The $N_-$ negatively charged matter fields $\tilde Q_{\tilde i}$ must identically vanish \vanishingii\ in a BPS configuration, so they do not yield bosonic zero modes. 

Now consider the Fermi zero modes of the $q_J=1$ BPS vortex.  Again, the counting is independent of the $\rho _i$ in \bozos\ (though $\rho _i\neq 0$ does dramatically affect the shape of the solutions) so we set $\rho _i=0$ for simplicity. As discussed after \qivortex, the Fermi zero mode equations \fermizmi\ and \mattzm\ then decouple among the matter flavors.  The photino and $\psi _{j=1}$ matter Fermion have the same solution as the minimal matter $(N_+, N_-)=(1,0)$  theory, giving  the normalizable, complex Fermi zero mode, $\Psi _1 \sim Q_-Q_1^{vortex}$. The remaining Fermion zero modes solve \mattzmx, i.e.
\eqn\fermizmelse{\pmatrix{0&D_{\bar z}\cr D_z &i A_0}\pmatrix{\psi _{j\uparrow}\cr \psi _{j\downarrow}}=0, \qquad \hbox{and}\qquad \pmatrix{0&D_{\bar z}\cr D_z &-i A_0}\pmatrix{\tilde \psi _{\tilde j \uparrow}\cr \tilde \psi _{\tilde j\downarrow}}=0,}
with $j=2\dots N_+$, and $\tilde j=1\dots N_-$.  For each such $j$ and $\tilde j$, \fermizmelse\  has one zero mode solution, with spin $\half \sign (n_i)$. 
As we have argued, for counting solutions and spins, we can replace \fermizmelse\ with the simpler version \baby, whose solutions are as in \vanishingf, \fplusis\ and \fminusis:
\eqn\fermielses{\qquad {\psi _{j>1, \uparrow}\over Q_1(z, \bar z)}= {\bar u_j \over \bar z-\bar z_0}, \qquad \hbox{and}\qquad {\tilde \psi _{\tilde j, \downarrow}\over Q_1^\dagger(z, \bar z)}= {d_{\tilde j}\over z-z_0}.}
The spinors $\bar u_{j>1}$ and $d_{\tilde j}$ give $N_++N_--1$  Fermi zero modes $\Psi _{j>1}$ and $\Psi _{\tilde j}$; all are non-normalizable, since all $\lim _{|z|\to \infty} |\psi|\sim 1/|z|$ in \fermielses.    The charges are as in \fermigen:
\eqn\fermisc{
\vbox{\offinterlineskip\tabskip=0pt
\halign{\strut\vrule#
&~$#$~\hfil\vrule
&~$#$~\hfil\vrule
&~$#$~\hfil\vrule
&~$#$~\hfil\vrule
&~$#$~\hfil\vrule
&~$#$~\hfil\vrule
&~$#$~\hfil
&\vrule#
\cr
\noalign{\hrule}
& & U(1)_{\rm spin} & SU(N_+) \times SU(N_-) & U(1)_A & U(1)_R & U(1)_J \cr
\noalign{\hrule}
&  \Psi_i         &\; \half  & \; \qquad (N_+ , 1)&\; 1  &\; -1 &\; 0 \cr
&  \Psi_{\tilde i}        &\; -\half  &\;\qquad (1, N_-) &\; 1  &\; -1 &\;0 \cr
&\prod _A\Psi _A &\;  \half \Delta N &\; \qquad (1,1)  &\;  N_{tot}&\; -N_{tot} &\; 0\cr
&|\Omega _\pm\rangle _{q_J=1}  &\;  -(k\pm \half \Delta N) &\; \qquad (1,1)  &\;  \mp \half N_{tot}&\; \pm \half N_{tot} &\; 1\cr
}\hrule}}
with $\Delta N\equiv N_+-N_-$ and $N_{tot}\equiv N_++N_-$. To save space, we here formally lump together $\Psi _1$ and $\Psi _{i>1}$, even though they are different, e.g. $\Psi _1$ is normalizable and $\Psi _{j>1}$ are not normalizable, consistent with the fact that $SU(N_+)$ is broken by \qivac\ to $SU(N_+-1)\times U(1)$.   

As discussed  \aspects\ and section 3.4, we quantize all $N_++N_-$ Fermi zero modes, including the non-normalizable ones.  This leads to a tower of $2^{N_++N_-}$ vortex states, with the top and bottom states $|\Omega _\pm\rangle _{q_J=1}$, with quantum numbers as in \fermisc.  The normalizable zero mode, $\Psi _1$, is identified with $Q_-$, so the states form $2^{N_++N_--1}$ BPS doublets \bpsp.  These come from quantizing the non-normalizable $\Psi _{j>1}$ and $\Psi _{\tilde i}$ Fermi zero modes:
\eqn\stateseg{ |a_{p, \tilde p}\rangle \equiv [\Psi _{j>1}]^p[\Psi _{\tilde i}]^{\tilde p}|\Omega _+\rangle, \qquad |b_{p, \tilde p}\rangle =Q_-|a_{p, \tilde p}\rangle}
meaning to fully antisymmetrize in $p$ different $\Psi _{j>1}$ and $\tilde p$ different $\Psi _{\tilde i}$, with $p=0\dots N_+-1$ and $\tilde p=0\dots N_-$. 
The states \stateseg\ all have $q_J=1$, with other quantum numbers
\eqn\stateegq{
\vbox{\offinterlineskip\tabskip=0pt
\halign{\strut\vrule#
&~$#$~\hfil\vrule
&~$#$~\hfil\vrule
&~$#$~\hfil\vrule
&~$#$~\hfil\vrule
&~$#$~\hfil\vrule
&~$#$~\hfil\vrule
&~$#$~\hfil
&\vrule#
\cr
\noalign{\hrule}
& & U(1)_{\rm spin} & SU(N_+-1) & SU(N_-) & U(1)_R \cr
\noalign{\hrule}
&  |a_{p, \tilde p}\rangle         &\; \half (p-\tilde p-\Delta N)-k& \; \pmatrix{N_+-1 \cr p}  & \pmatrix{N_-\cr \tilde p}  &\; -(p+\tilde p)+\half N_{tot}  \cr
&  |b_{p, \tilde p}\rangle         &\; \half (p+1-\tilde p-\Delta N)-k& \; \pmatrix{N_+-1 \cr p}  & \pmatrix{N_-\cr \tilde p}  &\; -(p+1+\tilde p)+\half N_{tot}  \cr
}\hrule}}
The omitted $U(1)_{\rm gauge}$ charge is screened by $Q_1^{vac}\neq 0$,  and we omit $U(1)_A$.   The $SU$ flavor singlets are $(p, \tilde p)=(0,0), \ (N_+-1, 0), \ (0, N_-), \ (N_+-1, N_-)$, with $|\Omega _-\rangle = |b_{N_+-1, N_-}\rangle$.    

If $k=\mp k_c\equiv \mp \half \Delta N$, the $X_\pm$ Coulomb branch exists, and  $|\Omega _\pm\rangle$ has spin $0$, and is an $SU(N_+-1)\times SU(N_-)$ singlet, consistent with \omegax\ and interpreting $X_\pm$  as a condensate of these states.    (By choice of $k$,  other $SU(N_+-1)\times SU(N_-)$ singlets states can have spin $0$; e.g. $|a_{0, N_-}\rangle$ has spin $0$ if $k=-\half N_+$, for all $N_-$.)

Consider the $(N_+, N_-)=(1,1)$ theory, i.e. $N_f=1$ SQED, with $k\in \Z$.  Taking $\zeta >0$, there are BPS vortices in the $M=Q\tilde Q=0$ vacuum \qivac. The $q_J=1$ BPS vortex has two Fermi zero modes (plus complex conjugates): the $\Psi _1\sim Q_+Q_1^{vortex}$ zero mode has spin $+\half$ and is normalizable, and the $\Psi _{\tilde 1}\equiv \Psi _2$ zero mode has spin $-\half$ and is not normalizable. Quantizing $\Psi _1$ and $\Psi _2$ \fermiquant\ gives two  BPS doublets:
 \eqn\vortexnfone{q_J=1:\qquad\pmatrix{|\Omega _+\rangle \cr Q_+|\Omega _+\rangle}, \qquad \pmatrix{ \Psi _{\tilde 1}|\Omega _+\rangle\cr Q_+\Psi _{\tilde 1}|\Omega _+\rangle,}}
with $\Psi _{\tilde 1}|\Omega _+\rangle \sim \bar Q_-|\Omega _-\rangle$ and $|\Omega _-\rangle \sim  Q_+\Psi _{\tilde 1}|\Omega _+\rangle$.  Since $\Psi _1 \sim Q_+Q_1^{vortex}$, $Q_+$ in \vortexnfone\ has the charges of $\Psi _1 Q_1^*$. The charges of the states are then, as in  \fermigen\ and \spectrum\ 
\eqn\nficharges{
\vbox{\offinterlineskip\tabskip=0pt
\halign{\strut\vrule#
&~$#$~\hfil\vrule
&~$#$~\hfil\vrule
&~$#$~\hfil\vrule
&~$#$~\hfil\vrule
&~$#$~\hfil\vrule
&~$#$~\hfil
&\vrule#
\cr
\noalign{\hrule}
&  &U(1)_{spin} &   U(1)_A & U(1)_J& U(1)_R  \cr
\noalign{\hrule}
&  |\Omega _+\rangle _{q_J=1}      & \; - \half k   &\; -1 &\; 1    &\; 1 \cr
&  Q_+|\Omega _+\rangle   & \;  -\half k +\half  &\; -1    & \; 1  & \; 0\cr
\noalign{\hrule}
}}}
\eqn\nfichargesii{
\vbox{\offinterlineskip\tabskip=0pt
\halign{\strut\vrule#
&~$#$~\hfil\vrule
&~$#$~\hfil\vrule
&~$#$~\hfil\vrule
&~$#$~\hfil\vrule
&~$#$~\hfil\vrule
&~$#$~\hfil
&\vrule#
\cr
\noalign{\hrule}
&  &U(1)_{spin} &   U(1)_A & U(1)_J& U(1)_R  \cr
\noalign{\hrule}
&  \bar Q_- |\Omega _-\rangle       & \; - \half k  -\half  &\; 1 &\; 1    &\; 0 \cr
&  |\Omega _-\rangle _{q_J=1}  & \;  -\half k &\; 1    & \; 1  & \; -1\cr
\noalign{\hrule}
}}}

The two BPS doublets in \nficharges\ and \nfichargesii\ reside in different Hilbert spaces, since they are connected via the non-normalizable $\Psi _2$ Fermi zero mode from $\psi _{\tilde Q}$.   For $k=0$, both $|\Omega _\pm \rangle _{q_J=1}$ have spin $0$, and quantum numbers consistent with \omegax: $|\Omega _+\rangle _{q_J=1}\sim X_+|0\rangle$ and $|\Omega _-\rangle _{q_J=1}\sim X_-^\dagger |0\rangle$.  We interpret $|\Omega _\pm\rangle$ in different Hilbert spaces as corresponding to $X_+X_-\sim 0$ in the chiral ring, and the disconnected $X_\pm$ branches of the $\zeta =0$ theory\foot{Parity is a symmetry for $k=0$ and maps  $X_+\leftrightarrow X_-$.   We can turn on a ($P$ odd) real mass $m_Q$ for $Q$ and $\tilde Q$ and then there is only one Coulomb branch, $X_\pm$ if $m(X_\pm)=-m_Q\pm \zeta=0$; $m_Q\neq 0$ also eliminates the non-normalizable $\psi _{\tilde Q}$ zero mode.  There is then a BPS state matching either $X_+|0\rangle$, or $X_-^\dagger |0\rangle$, depending on $\sign (m_Q\zeta)$.  Taking $m_Q\to 0$ requires both doublets in \vortexnfone.}. 

The $W=MX_+X_-$ dual \AharonyBX\ must have the same structure: the map from the  $X_+|0\rangle$ to the  $X_-^\dagger |0\rangle$ BPS state must involve (in addition to the normalizable $Q_+$ zero mode), a $\sim 1/|z|$ non-normalizable $\psi _M=Q\tilde \psi_{\tilde Q}$ zero mode.   Again, we propose that this reflects that the map from $X_- ^\dagger |0\rangle$ to $X_+ |0\rangle$ is via $X_+ X_-\sim \overline {F_M}\sim  \{ \bar Q^\alpha ,[ \bar Q_\alpha, \bar M ]\}\sim 0$ in the chiral ring.   This tentative interpretation should be further clarified, perhaps in future work.  

\newsec{Cases with $Q_i^{vac}\neq 0$ for matter with $n_i\neq 1$.}

If a matter field $Q_1$, with $n_1>1$, has an expectation value \qivac\
(negative $n_1$ can be obtained via charge conjugation of the present discussion), $Q_1^{vac}\neq 0$  breaks $U(1)_{\rm gauge}\to \Z_{n_1}$, a discrete gauge symmetry, a.k.a. a $\Z _{n_1}$ orbifold.  See \BanksZN, and references cited therein, for more about $\Z _{n_1}$ gauge theory. 
Before the $\Z _{n_1}$ orbifold projection, the Fermion zero modes are essentially the same as in section 4, with $|n_i|$ Fermion zero modes $\Psi _{i=1, p=1\dots |n_i|}$ for each matter field $Q_i$, and charges as in \fermigen.  This includes $n_1$ Fermi zero modes (one is the supercharge) coming from matter field $Q_1$ and the photino, from eqns. 
\fermizmi, \mattzm.  The Fermi zero modes are quantized as in \fermiquant, giving a tower of $2^{\sum _i |n_i|}$ states, and one then projections to $\Z _{n_1}$ gauge invariant states.  The top and bottom states $|\Omega _\pm\rangle _{q_J=1}$ \omegapm\ survive the $\Z _{n_1}$ projection, with quantum numbers again matching with $X_+$ and $X_-^\dagger$.

As a special case, recall from \aspects\ that if the charges all have a common integer factor, $n_i=n\tilde n_i$, with $n$ and $\tilde n_i$ integer, the theory is simply a $\Z _n$ orbifold of a rescaled theory:
\eqn\rescaling{n_i\to  \tilde n_i\equiv n_i/n, \qquad q_J\to \tilde q_J=nq_J, \qquad \tilde a\to a/n.}
Note that $q_J\in \Z$, while $\tilde q_J\in n\Z$, and $a$ has periodicity $a\sim a+2\pi$, while $a\sim a+2\pi/n$.    Consider e.g. the theory of a single matter field, $Q_1$, with charge $n_1>1$, which is equivalent 
to a $\Z _{n_1}$ orbifold of the rescaled theory with matter of charge $\tilde n_1=1$.  Since the $q_J=1$ vortex of the original theory maps  \rescaling\ to a $\tilde q_J=n_1$ vortex of the rescaled theory, it has    $n_1$ complex Bosonic zero modes (the locations $z_1, \dots, z_{n_1}$ of the individual vortex cores in the rescaled theory), and $n_1$ Fermionic zero modes, $\Psi _1, \dots, \Psi _{n_1}$, prior to the $\Z _{n_1}$ orbifolding.  Quantizing the $\Psi _{A=1\dots n_1}$ as in \fermiquant, gives a tower of  $2^{n_1}$ states.  The top and bottom states,  $|\Omega _\pm\rangle _{q_J=1}$, have charges as given by \omegapmx\ and  \fermigen, here with $k_c=\half n_1^2$.  These states are $\Z _{n_1}$ invariant, and their charges match those of  $X_+$ and $X_-^\dagger$ in \onengen.  For $k=\mp k_c$, the operator $X=X_\pm$ is $U(1)_{gauge}$ neutral, with spin 0, and labels a half-Coulomb branch. This theory is a 
$\Z_{n_1}$ orbifold of a free field theory \aspects, with $X^{1/n_1}$ the free field. 

We can also consider BPS vortices in vacua with $Q_i^{vac}\neq 0$ for multiple fields, of different charges $n_i$, with all $\sign (n_i)=\sign (\zeta)$ \vanishing , i.e. a weighted projective space, with weights $n_i$.   The Fermi zero mode analysis for the general case is then complicated by the couplings among flavors in \mattzm.  In any case, the counting and charges of the Fermi zero modes cannot be affected by continuous moduli, so they must again be as as \fermigen.   

In conclusion, in all cases the BPS vortex states $|\Omega _\pm \rangle _{q_J=1}$ have quantum numbers compatible with \omegax.  For $k=\mp k_c$, 
it is a spin $0$ BPS state, which becomes massless for $\zeta \to 0$ and can condense to give a dual Higgs description of the $X_\pm$ Coulomb branch.

\bigskip
\noindent {\bf Acknowledgments:}

I would especially like to thank Nathan Seiberg for many illuminating discussions, key observations, and helpful suggestions.     I would also like to thank Juan Maldacena, Ilarion Melnakov,  Silviu Pufu, Sav Sethi, and David Tong, for useful discussions or correspondence.  I would like to thank the organizers and participants of the workshops {\it String Geometry and Beyond} at the Soltis Center, Costa Rica, and the KITP program {\it New Methods in Nonperturbative Quantum Field Theory} for the opportunities to discuss this work, and for many stimulating discussions.  I would especially like to thank the KITP, Santa Barbara, for hospitality and support in the final stage of this work, in part funded by the National Science Foundation under Grant No. NSF PHY11-25915. This work was also supported by the US Department of Energy under UCSD's contract DE-SC0009919, and the Dan Broida Chair.

\appendix{A}{Additional details, conventions, and notation}

In components, the lagrangian \lclass\ is 
\eqn\lclasscont{\eqalign{\CL _{cl}&=-{1\over 4e^2} F_{\mu \nu}F^{\mu \nu}+{k\over 4\pi}\epsilon ^{\mu \nu\rho}A_\mu \partial _\nu A_\rho -{1\over 2e^2} (\partial \sigma )^2- \sum _i |D^{(n_j)}_\mu Q _i |^2 -V_{cl}\cr 
&-\bar \lambda ({i\over e^2}\slashchar{\partial} +{k\over 4\pi})\lambda  -\sum _j \bar \psi _j (i\slashchar{D}^{(n_j)} + m_j(\sigma)) \psi _j +i\sqrt{2}\sum _j n_j(Q _j^*\lambda \psi _j -
 Q _j\bar \lambda \bar \psi _j).}}
We use \WessCP\  conventions\foot{So $(-++)$ signature and $
\gamma^{\mu=0,1,2}_{\alpha\beta}= (-{\bf 1}, \sigma^1, \sigma^3),$ i.e. $(\gamma ^\mu)_\alpha {}^\beta = \gamma ^\mu _{\alpha \lambda}\epsilon ^{\lambda \beta}= (-i\sigma ^2, -\sigma _3, -\sigma _1)$.} (reduced from 4d to 3d along the $x^{\mu =2}$ direction, see \DumitrescuIU), though this introduces an unfortunate, non-standard sign convention\foot{Since 
\WessCP\ uses $D_\mu ^{(n_j)}\equiv \partial _\mu +i n_j A^{W\& B}_\mu$ in mostly plus signature, $A^{W\& B}_\mu = - A_\mu ^{usual}$; this is also apparent from their $\CL \subset -j^\mu A_\mu ^{W\& B}$.  Consequently, $[D^{(n_j)}_1, D^{(n_j)}_2]=i n_j F_{12}^{W\& B}=-i n_j F_{12}^{(usual)}$, which changes the names of BPS vs anti-BPS with respect to much of the vortex literature. This could be fixed by introducing a minus sign in the definition \qjis\ of $q_J$,  but that introduces sign differences with other literature, e.g. the  definitions of $X_\pm$ in \refs{\AharonyBX, \aspects}, so we will not do that here.} for the gauge field.  The gauge supermultiplet fields $(A_\mu, \lambda, \sigma)$ have 4d mass dimensions, e.g. $[\lambda ]=3/2$, with $[e^2]=1$, while $[Q _i]=1/2$ and $[\psi _i]=1$ for the matter.  The scalar potential $V_{cl}$ in \lclasscont\ is 
\eqn\vcl{V_{cl}={e^2\over 32 \pi ^2} \left( \sum_i 2\pi n_i |Q _i|^2 -\zeta -k \sigma\right)^2 + \sum_i (m_i + n_i\sigma)^2 |Q _i|^2~~. }
In a configuration where the fields asymptote to a zero of \vcl, the total energy of \lclasscont\ (with $m_i=0$) can be written (using \identity\ and \gaussis), as  (with $F_{12}\equiv F_{12}^{W\& B}$)
  \eqn\etwo{\eqalign{E&=\pm \zeta q_J+{1\over 2e^2}\int d^2 x \left(
(F_{12}\mp D)^2+(F_{i0}\mp \partial _i \sigma)^2+(\partial _0\sigma )^2\right)\cr
&+\int d^2 x \sum _j \left(|(D_0\mp i n_j\sigma)Q _j \sigma |^2+|(D_1\mp i D_2)^{(n_j)}Q _j|^2\right) \geq \pm \zeta q_J,}}  
with $D$ as in \dis.  The BPS (resp. anti-BPS) configurations saturates the inequality for upper (resp. lower) sign choice and $\zeta q_J>0$ (resp. $\zeta q_J<0$).

\listrefs
\end